\documentclass{aastex631}

\usepackage{enumitem}
\usepackage{xcolor}

\received{Aug 27, 2024}
\revised{Dec 07, 2024}
\accepted{Feb 08, 2025}

\shorttitle{Quasars behind the Galactic Plane from LAMOST DR10}
\shortauthors{Z. Y. Huo et al.}

\graphicspath{{./}{figures/}}

\begin{document}

\title{Finding Quasars Behind the Galactic Plane: Spectroscopic Identifications of  $\sim$1300 New Quasars
at $|b| \le 20\degr$ from LAMOST DR10}


\author[0009-0003-3066-2830]{Zhi-Ying Huo}
\affil{National Astronomical Observatories, Chinese Academy of Sciences, Beijing 100101, People's Republic of China}

\author[0000-0002-0759-0504]{Yuming Fu}
\affil{Leiden Observatory, Leiden University, Einsteinweg 55, 2333 CC Leiden, The Netherlands}
\affil{Kapteyn Astronomical Institute, University of Groningen, P.O. Box 800, 9700 AV Groningen, The Netherlands}

\author[0000-0003-3250-2876]{Yang Huang}
\affil{School of Astronomy and Space Science, University of Chinese Academy of Sciences, Beijing 100049, People's Republic of China}

\author[0000-0003-2471-2363]{Haibo Yuan}
\affil{Department of Astronomy, Beijing Normal University No.19, Xinjiekouwai St, Haidian District, Beijing, 100875, P.R. China}
\affil{Institute for Frontiers in Astronomy and Astrophysics, Beijing Normal University, Beijing 102206, People's Republic of China}

\author[0000-0002-7350-6913]{Xue-Bing Wu}
\affil{Department of Astronomy, School of Physics, Peking University, Beijing 100871, People's Republic of China}
\affil{Kavli Institute for Astronomy and Astrophysics, Peking University, Beijing 100871, People's Republic of China}
\affil{National Astronomical Observatories, Chinese Academy of Sciences, Beijing 100101, People's Republic of China}

\author{Maosheng Xiang}
\affil{National Astronomical Observatories, Chinese Academy of Sciences, Beijing 100101, People's Republic of China}

\author{Xiao-Wei Liu}
\affil{South-Western Institute For Astronomy Research, Yunnan University, Kunming 650500, People's Republic of China}

\author[0000-0001-8879-368X]{Bing Lyu}
\affil{Kavli Institute for Astronomy and Astrophysics, Peking University, Beijing 100871, People's Republic of China}

\author[0009-0006-7193-4443]{Hao Wu}
\affiliation{Department of Astronomy, School of Physics, Peking University, Beijing 100871, People's Republic of China}
\affiliation{Kavli Institute for Astronomy and Astrophysics, Peking University, Beijing 100871,  People's Republic of China}

\author{Jian Li}
\affil{Beijing Planetarium, Beijing Academy of Science and Technology, Beijing, 100044, People's Republic of China}

\author[0000-0002-6610-5265]{Yanxia Zhang}
\affil{National Astronomical Observatories, Chinese Academy of Sciences, Beijing 100101, People's Republic of China}

\author{Yanli Ai}
\affil{College of Engineering Physics, Shenzhen Technology University, Shenzhen 518118, People's Republic of China}
\affil{Center for Intense Laser Application Technology, Shenzhen 518118, People's Republic of China}

\author[0000-0002-8402-3722]{Junjie Jin}
\affil{National Astronomical Observatories, Chinese Academy of Sciences, Beijing 100101, People's Republic of China}

\email{zhiyinghuo@nao.cas.cn}

\begin{abstract}

Quasars behind the Galactic plane (GPQs) are excellent tracers to probe the chemistry and kinematics of the 
interstellar/intergalactic medium (ISM/IGM) of the Milky Way along sight lines via absorption line spectroscopy. 
Moreover, the quasars located at low Galactic latitudes will fill the gap in the spatial distribution of known quasars 
near the Galactic plane, and can be used to construct an astrometric reference frame for accurate measurements of 
proper motions (PMs) of stars, and substructures of the Milky Way.
We started a survey of background quasars in the low Galactic latitude region since the LAMOST phase II survey in 2017.
Quasar candidates have been selected from the optical and infrared photometric data of Pan-STARRS1 and WISE surveys
based on their variability and color properties.
In this paper, we present a sample of 1982 spectroscopically confirmed GPQs with $|b| \le 20\degr$ based on 
LAMOST Data Release 10 (DR10). Among them, 1338 are newly discovered. 
Most GPQs are located around $240\degr<l<90\degr$, and the spatial distributions are non-uniform.
These GPQs have a magnitude distribution with a peak at $i$-mag 19.0, and mostly 
around 18.0-19.5\,mag. The peak of redshift distributions is around $\sim$1.5, and most GPQs have
redshifts between 0.3 and 2.5. 
Our finding demonstrates the potential discovery space for the GPQs from the spectroscopic surveys and 
the promising applications for future research.

\end{abstract}


\section{Introduction} \label{sec:intro}

Since the discovery of the first quasar 3C273 with a redshift of $z = 0.158$ \citep{1963Natur.197.1040S}, astronomers have dedicated significant efforts to the search for quasars. Such searches aim to study various astrophysical phenomena, including the large-scale structures of the universe \citep[e.g.,][]{2005MNRAS.356..415C, 2019A&A...629A..86B}, the central supermassive black holes \citep[e.g.,][]{2008ApJ...680..169S, 2024NatAs...8..126B}, the reionization processes \citep[e.g.,][]{2006AJ....132..117F, 2019MNRAS.484.5094G}, 
and the formation and evolution of galaxies \citep[e.g.,][]{1998A&A...331L...1S, 2024Natur.627..281A}.

The Sloan Digital Sky Survey \citep[SDSS;][]{2000AJ....120.1579Y} has contributed the largest sample of quasars to date
\citep[e.g.,][]{2010AJ....139.2360S, 2017A&A...597A..79P, 2018A&A...613A..51P, 2020ApJS..250....8L}.
Until the recent 16th data release of SDSS quasar catalog (DR16Q), the number of spectroscopically confirmed
quasars has reached $\sim$ 750,000 \citep{2020ApJS..250....8L}.
Most SDSS quasars are distributed in the North Galactic Cap with an area of $\sim$ 7500\,deg$^2$, and the South Galactic Cap with an area of $\sim$ 3100\,deg$^2$ \citep[see][]{2011ApJS..193...29A, 2012ApJS..203...21A}. 
Near the Galactic plane, especially at $|b| \le 20\degr$, the number of spectroscopically identified quasars
is extremely few, due to the lack of systematic spectroscopic surveys of quasars in 
the low Galactic latitudes, where there is usually high dust extinction \citep[see][]{1998ApJ...500..525S, 
2016ApJ...821...78S, Planck_2016} and crowded stellar fields \citep[e.g.][]{2006AJ....131.1163S, 2009A&A...495..819R}.

Nevertheless, quasars behind the Galactic plane (GPQs) have important astrophysical applications.
Firstly, quasars are ideal sources for building celestial reference frames (CRFs) because of their small proper motions (PMs)
and parallaxes \citep[e.g.][]{2018A&A...616A..17A, 2023A&A...674A..32B, 2018A&A...616A..14G, 2022A&A...667A.148G}. 
A larger number of GPQs can help improve the CRFs by filling the gap in spatial distributions of quasars in the area of the Galactic plane. 
The current sparse distributions of background quasars behind the Galactic plane can be seen from the recently adopted CRFs by Gaia DR2 and DR3  \citep[][]{2018A&A...616A..14G, 2022A&A...667A.148G}.
Precise PM measurements enabled by exemplary CRFs are fundamental for many astrophysical problems, 
including the dynamics of stars and substructures, and the formation and evolution of the Milky Way and the Local Group
\citep[][]{2021A&A...649A...9G, 2021ARA&A..59...59B}.

Secondly, based on absorption-line spectroscopy, bright GPQs are excellent tracers for probing the foreground 
interstellar/intergalactic medium (ISM/IGM) of the Milky Way along the lines of sight. 
This enables the investigation of the physical properties of the ISM/IGM, 
such as the kinematics, the compositions and the spatial distributions. 
Based on Hubble Space Telescope Quasar Absorption Key Project, \citet[][]{1993ApJ...413..116S, 2000ApJS..129..563S} 
revealed that the Milky Way halo gas was undergoing a great complexity of chemical compositions, ionization states, and kinematics.
Using similar techniques, \citet[][]{2008A&A...487..583B, 2012A&A...542A.110B} systematically analyzed optical absorption line
data of Ca~\textsc{ii} and Na~\textsc{i} with follow-up HI 21-cm emission line observations of hundreds of QSOs.  
\citet[][]{2015MNRAS.452..511M} mapped out Ca~\textsc{ii} and Na~\textsc{i} absorption induced by ISM/IGM of the Milky Way covering more than
9000\,deg$^2$ with about 300,000 extragalactic spectra from the SDSS.
QSO absorption-line measurements of the ISM/IGM of the Milky Way indicate complex processes of galaxy formation and evolution,
including gas accretion from the IGM, merger events and outflow of gaseous material from galactic winds 
\citep[see][for a review]{2006RvMA...19...31R}.

Another application of GPQs is adaptive optics observation of quasars and their host galaxies. For adaptive optics, natural guide star
should be within about dozens of arcseconds of the target, this is easier to achieve in the area of the  Galactic plane than outside.
\citet[][]{2005A&A...434..469F} reported on the Very Large Telescope of European Southern Observatory (ESO-VLT)
 near-IR adaptive optics imaging of three QSO host galaxies 
and their environments at z$ \sim$ 2.5, observations were carried out in the condition that a star brighter than $V = 14$ was found within 30 arcseconds. 
 \citet[][]{2016MNRAS.458....2R} presented an imaging observation campaign conducted with the Subaru Telescope
 adaptive optics system on 28 gravitationally lensed quasars and candidates (all targets must be close to a star of
 suitable brightness to perform adaptive optics correction), provided accurate astrometry and relative photometry, and 
 derived robust morphology for the lensing galaxies and the quasar host galaxies at redshift $\sim$ 2.  
 \citet[][]{2024Natur.627..281A} reported a dynamical measurement of the mass of the black hole in quasar J092034.17+065718.0 at redshift
 2.30 with GRAVITY+ at the Very Large Telescope Interferometer using the new wide-field, off-axis fringe-tracking mode,
 by spatially resolving the broad-line region.
 A star with a $K$-band magnitude of 10.4 and a distance of 12\farcs7 away is used as the fringe-tracking object. 
 %

A number of studies have been carried on to search for background quasars behind nearby galaxies, with the purpose
to study their PMs and ISM/IGM. 
Thousands of AGNs/QSOs and candidates have been identified behind the Magellanic Clouds 
\citep[e.g.,][and references therein]{2009ApJ...701..508K, 2012ApJ...747..107K, 2011ApJS..194...22K, 2012ApJ...746...27K, 2013ApJ...775...92K}.  
\citet[][]{2010RAA....10..612H, 2013AJ....145..159H, 2015RAA....15.1438H}  presented $\sim$2000 background quasars in the 
vicinity of the Andromeda (M31) and Triangulum (M33) galaxies. 

Similarly, a few studies have been dedicated to finding quasars behind the Galactic plane, each presenting a few tens of new spectroscopically identified GPQs. \citet[][]{2007ApJ...664...64I} reported the discovery of 40 bright QSOs/AGNs at low Galactic 
latitude ($|b| < 20\degr$) that were selected from radio and near-infrared (NIR) data.
\citet[][]{2017RAA....17...32H} presented a sample of 80 new quasars discovered in the area of the Galactic Anti-Center
with $|b| \le 30 \degr$ based on LAMOST DR3, all these quasars were discovered serendipitously by LAMOST, which were originally
targeted as stars with bluer colors, or variability. \citet[][]{2024ApJS..273...21W} reported the spectroscopic identifications 
of 72 UV-bright AGNs at $|b|<30\degr$, among which 25 are new discoveries. 
So far, only 14,279 QSOs/AGNs or candidates at $|b|<20\degr$ have been included in the Million Quasars Catalog 
\citep[Milliquas v8;][]{2023OJAp....6E..49F}.

To boost the efficiency of systematic searches for GPQs, \citet[][]{2021ApJS..254....6F} developed a candidate 
selection method based on transfer learning, providing a catalog of $\sim$ 160,000 GPQ candidates within $|b|<20\degr$ 
using Pan-STARRS1 \citep[PS1;][]{2016arXiv161205560C} DR1 and the final catalog release of the 
Wide-field Infrared Survey Explorer \citep[AllWISE;][]{2010AJ....140.1868W, 2013wise.rept....1C}. 
Of these 160,000 quasar candidates, 30,169 have Dec$\ge -10\degr$ and $i$-mag $\le$19.5,
which are suitable for LAMOST survey observations.
\citet[][]{2022ApJS..261...32F} presented early spectroscopic observations of 230 GPQ candidates from 
\citet[][]{2021ApJS..254....6F} since 2018 with several optical telescopes, including the Xinglong 2.16\,m Telescope, 
the Lijiang 2.4\,m Telescope, the 200\,inch Hale Telescope, the ANU 2.3\,m Telescope and the McGraw-Hill 1.3\,m Telescope. This spectroscopic campaign has confirmed 204 GPQs, among which 191 are newly discovered, and the successful identification rate of GPQs is 85\%.

In this paper, we will report our pilot work of the LAMOST GPQ program, in which $\sim$1300 new GPQs are identified from LAMOST DR10. Most of these new GPQs have been observed as GPQ candidates during the LAMOST phase II survey, from September 2017 to June 2022. 
Quasars discovered serendipitously within $|b|\leq 20\degr$ in LAMOST DR10 are also included in this paper.
During the preparatory stage of the LAMOST phase II survey, we proposed to fill the gap in the spatial distribution of known
quasars in the area of the Galactic plane. At that time, there were no appropriate quasar candidates for the Galactic plane area.
We therefore adopted the PS1 quasar candidate catalog from \citet[][]{2016ApJ...817...73H} selected based on 
the variability. 
In 2021, with the publication of a dedicated catalog of quasar candidates behind the Galactic plane by \citet[][]{2021ApJS..254....6F}, we expanded the list of GPQ candidates to include the newly available GPQ candidates from this work as well.

The GPQ survey is currently continuing during the LAMOST phase III survey (2023-2027), and the GPQ candidates for the LAMOST phase III survey
are the combination of quasar candidates from \citet[][]{2016ApJ...817...73H} and \citet[][]{2021ApJS..254....6F}.
With an $i$-mag limit of 19.5 and a cut on Gaia PM, $\sim$48,000 quasar candidates with Dec $\ge -10\degr$ are submitted (Table \ref{tab:gpq_cand}). 
In the fiber allocation procedure, the GPQ candidates have the highest priority. 
However, GPQ candidates are allocated in the observation plates of medium brightness, which need to be observed under 
particularly good weather conditions. In addition, LAMOST is now spending half of the time doing the medium-resolution ($R \sim 7500$) time domain surveys on stars,
which reduces the time for low-resolution surveys significantly. 
With the newly approved scientific observational proposals, LAMOST phase III is initiating a new round of sky survey 
coverage (Xu et al. 2025, in preparation).
Considering both the previous sky survey progress and the current improved performance of LAMOST, 
we made a conservative estimate that about 5000 GPQs are expected to be discovered in the LAMOST phase III survey.

The paper is organized as follows. In Section 2, we elucidate the origins of the GPQ candidates.
The observation details and data reductions are described in Section 3. 
The GPQ catalog is presented in Section 4.
Finally, the properties of this GPQ sample are discussed in Section 5 and the summary is given in Section 6. 
We adopt a flat universe with $\Omega_M=0.3$ and $\Omega_{\Lambda}=0.7$, and the cosmology parameter
$H_0 = 70~\mathrm{km~s^{-1}~Mpc^{-1}}$.

\section{Candidate Selection} \label{sec:Candsel}

Here, we briefly review the two quasar candidate catalogs adopted during the LAMOST phase II survey, 
which are from \citet[][]{2016ApJ...817...73H} and \citet[][]{2021ApJS..254....6F}. 
The PS1 3$\pi$ survey has observed the entire sky north of Declination  $-30\degr$ typically seven times
in each of its five filter bands ($grizy_{\rm P1}$) over the first 3.5 years, and has enormous potential for all-sky identifications
of variable sources \citep[][]{2016arXiv161205560C}. 
\citet[][]{2016ApJ...817...73H} developed an approach for quantifying the statistical properties of non-simultaneous, 
sparse, multi-color light curves through light curve structure functions, turning PS1 into a $\sim$ 35-epoch survey.
Using a Random Forest Classifier, \citet[][]{2016ApJ...817...73H} identified a sample of $\sim$ 1 million QSO
candidates, as well as an unprecedentedly large and deep sample of $\sim$ 150,000 RR Lyrae candidates based on their
variability and mean PS1 and WISE colors. 
Far from the Galactic plane, the QSO and RR Lyrae samples can be selected with purity of $\sim$ 75\% and 
completeness of $\sim$ 92\%, with PS1 data on SDSS Stripe 82 as training set. 
\citet[][]{2016ApJ...817...73H} obtained a homogeneous distribution of the QSO candidates down to $|b| \sim 10 \degr$,
until dust extinction and disk star contamination become severe, which is shown in their Figures 9 and 18.
Around the Galactic plane, the number of GPQ candidates decreases significantly because of the high dust extinctions
and star contaminants, the estimation of completeness and purity is not provided. 
QSO candidates with probability thresholds of $p_{\rm QSO} \ge 0.2$ and PS1 error-weighted mean $i_{\rm P1}$ band magnitudes
brighter than 20.0 are selected for the LAMOST observations. This brings 63,421 QSO candidates with $|b| \le 20\degr$ in the PS1
footprint.
Please note that the magnitude limit for the initial GPQ candidates is restricted to $i$-mag $\le 20.0$ before 2021; 
however, the magnitude limit of the subsequent quasar candidates is restricted to $i$-mag $\le 19.5$ to ensure 
the signal to noise ratio (S/N) of the LAMOST spectra, including those from
\citet[][]{2021ApJS..254....6F} and the ongoing quasar candidates for the LAMOST phase III survey.
The information of the GPQ candidates adopted by LAMOST at different times is listed in Table \ref{tab:gpq_cand}.

Historically, the search for quasars has primarily been focused on high Galactic latitudes, while the Galactic plane has been regarded as the ``zone of avoidance" for quasar survey.
Until recently, \citet[][]{2021ApJS..254....6F} presented a systematic search for quasar candidates behind the Galactic plane.
They adopted a transfer-learning method to select GPQ candidates, which mocks the color-magnitude distributions of GPQs and addresses the class imbalance problem of star-galaxy-quasar classification. They applied the XGBoost algorithm to PS1 and AllWISE photometry to classify objects as stars, galaxies, and quasars, and used an additional cut on Gaia PM to remove stellar contaminants. 
As a result, they created a reliable GPQ candidate catalog with 160,946 sources located within $|b| \le 20\degr$ in the PS1-AllWISE footprint.
Photometric redshifts of GPQ candidates achieved with the XGBoost regression algorithm showed that the selection method of \citet{2021ApJS..254....6F} could identify quasars in a wide redshift range ($0 < z \le 5$). 
\citet[][]{2021ApJS..254....6F} used the Simbad database \citep[][]{2000A&AS..143....9W} 
and the Milliquas \citep[][]{2023OJAp....6E..49F} to validate their GPQ candidate sample, 
and concluded that the purity was as high as 90\%, 
and the completeness was higher than 95\%. The estimations of purity and completeness are heavily dependent on the validation set, 
and the true purity and completeness may vary at different locations in the Galactic plane.

Within $|b| \leq 20\degr$, 62\% of candidates in \citet[][]{2016ApJ...817...73H} coincide with those of \citet{2021ApJS..254....6F}, indicating the effectiveness of both candidate selection methods, and the advantage of combining both samples for higher completeness. 
For the LAMOST observations, the magnitude limit of GPQ candidates from \citet{2021ApJS..254....6F} is set to $i_{\rm P1}$ 
band 19.5\,mag considering the low S/Ns of the spectra of quasars within magnitude ranges of 19.5$\sim$20.0 obtained with LAMOST
in the previous work \citep[see][]{2016AJ....151...24A, 2018AJ....155..189D, 2019ApJS..240....6Y, 2023ApJS..265...25J}. 
This leads to 30,169 candidates that are suitable for LAMOST observations (Table \ref{tab:gpq_cand}).

In addition to the two quasar candidate catalogs described above, there are two supplement origins for GPQs in this work.
One origin is that the quasars discovered serendipitously from objects that are originally targeted by LAMOST as stars
with extremely blue or red colors, candidates targeted as variable, or young stellar objects (YSOs), 
for example the 151 (80 new) quasars in the Galactic Anti-Center area listed in \citet[][]{2017RAA....17...32H} based on LAMOST DR3.
The other origin is that some quasar candidates in the vicinity of M31 and M33, located at $|b| \le$ 20$\degr$,  
are included in this work \citep[see][for more details]{2010RAA....10..612H, 2013AJ....145..159H, 2015RAA....15.1438H}.

\begin{deluxetable*}{ccccc}[htb!]
\tablecaption{Information of the GPQ candidates adopted by LAMOST at different times. \label{tab:gpq_cand}}
\tablehead{\colhead{Catalog} & \colhead{$i$-mag limit} & \colhead{PM cut} & \colhead{Number}  & \colhead{Observation Phase}}
\startdata
\citet[]{2016ApJ...817...73H} & 20.0 & No  &63,421 &  2017-2022 \\
\citet[]{2021ApJS..254....6F} & 19.5 & Yes & 30,169 & 2021-2022 \\
The combination of \citet[]{2016ApJ...817...73H} and \citet[]{2021ApJS..254....6F} & 19.5 & Yes & 48,026 & 2023-2027 \\
\enddata
\end{deluxetable*}

\section{Observations and Data Reduction} \label{sec: Obs_red}

LAMOST, also called the Guoshoujing telescope, is a quasi-meridian reflective Schmidt telescope with an effective 
aperture that varies from 3.6\,m to 4.9\,m (depending on the declination) and an angular diameter of the field of view of 
is 5$\degr$ for $-10\degr \le \delta \le 60\degr$ and 3$\degr$ for $60\degr \le \delta \le 90\degr$
\citep[][]{1996ApOpt..35.5155W, 2004ChJAA...4....1S, 2012RAA....12.1197C}. 
LAMOST can simultaneously obtain 4000 spectra in a single exposure. 
The spectral resolution is $R\sim$1800 for the low-resolution mode, with a wavelength range of 3700 to 9000${\rm \AA}$, 
which is divided into blue (3700-5900${\rm \AA}$) and red (5700-9000${\rm \AA}$) spectroscopic channels.

This current work is based on LAMOST DR 10, which includes all data collected from October 2011 to June 2022.
Specifically, quasar candidates located in the Galactic plane have been targeted by LAMOST since the onset of the phase II survey in September 2017, which concluded in June 2022. 
Initially, until May 2021, the target list exclusively included quasar candidates from \citet[][]{2016ApJ...817...73H}. Starting in May 2021, the list was expanded to also incorporate GPQ candidates from \citet[][]{2021ApJS..254....6F}.

The raw CCD image data obtained from observations were reduced using the two-dimensional (2D) and one-dimensional (1D) pipelines of LAMOST \citep[see][for more details]{2015RAA....15.1095L}. The 2D pipeline performs several tasks to extract 1D spectra from the raw images, 
including bias subtraction, cosmic-ray removal, spectral tracing and extraction, wavelength calibration, flat fielding,  
sky subtraction and flux calibration. Subsequently, the 1D spectra are classified into three types: ``STAR'', ``GALAXY'' and ``QSO'', through template matching by 1D pipeline. It should be noted that the category ``UNKNOWN'' has been obsolete since DR9. 
Alongside classifications, radial velocities for stars and redshifts for galaxies and quasars are simultaneously determined. 
 
We selected all spectra of quasar candidates within $|b| \leq 20 \degr$ observed by LAMOST, including either those specifically 
selected as GPQ candidates (see Section \ref{sec:Candsel}), or those classified as ``QSO'' by the LAMOST 1D pipeline but not 
included in the GPQ candidates. 
In total, there are 3486 spectra corresponding to 3030 unique targets within this Galactic latitude range. 
Among them, 2454 are quasar candidates, and 576 are originally targeted as stars, YSOs, variable objects, 
WD or RR Lyrae candidates, but classified as ``QSO" by the 1D pipeline of LAMOST.

\section{Identifications and the Catalog} \label{sec:Iden_cat}

Quasars are relatively easy to identify given their characteristic broad and strong emission lines. Since a certain number of quasar
candidates are fainter than 19\,mag in $i$-band (between 19 and 20\,mag), their S/Ns are relatively low.
Consequently, we visually examined all these 3486 spectra by at least two co-authors.
With the help of a JAVA program ASERA \citep[][]{2013A&C.....3...65Y}, all those spectra with at least one or two emission lines
matching to the quasar template are identified as quasars. 
The redshift is measured simultaneously with at least one or two available typical quasar emission lines matching with the templates, 
such as Ly$\alpha$, C\,{\sc iv}, C\,{\sc iii}], Mg\,{\sc ii}, H$\beta$, [O\,{\sc iii}] and H$\alpha$.
For a few quasars near redshift of $\sim$1.0, only the Mg\,{\sc ii} line is detected, which falls in
the overlap range of the blue and red channels (5700-5900\AA), where the instrument throughputs are relatively low, 
and spike- or trough-like artifacts may appear in the flux-calibrated spectra. At the same time, other emission lines,
such as H$\delta$, H$\gamma$, and H$\beta$ are all moved to wavelengths longer than 7800\AA, where the sky emission lines seriously contaminate the LAMOST
spectra. Therefore quasars with redshifts of $\sim$1.0 may have only one
Mg\,{\sc ii} emission line identified.

At last, 1982 unique quasars are identified in the area of the Galactic plane with $|b| \le 20\degr$ from LAMOST DR10.
Among the 2454 observed quasar candidates, 1949/16/165/324 are classified as 
QSO/GALAXY/STAR/UNKNOWN relatively after visual examination, and the success rate of quasar identification is 79\%. 
Among the 576 targets originally targeted as stars, YSOs, variable targets, WD or RR Lyrae candidates,
33/6/177/360 are classified as QSO/GALAXY/STAR/UNKNOWN relatively. 
There are around $\sim$680 observed targets which have relatively low S/Ns, and we can not provide classifications for them.
Among the 1982 identified quasars, 1338 quasars are newly discovered after cross-matching the full sample with known 
quasar catalogs which contain quasars located at $|b| \le 20\degr$, including: 
\begin{itemize}[noitemsep]
     \item[(1)] The Million Quasars Catalogue, v8 \citep[Milliquas;][]{2023OJAp....6E..49F}.
     \item[(2)] Quasars in the Galactic Anti-Center Area from LAMOST DR3 \citep[][]{2017RAA....17...32H}.
     \item[(3)] Finding Quasars behind the Galactic plane. II. Spectroscopic Identifications of 204 Quasars at $|b|< 20 \degr$ \citep[][]{2022ApJS..261...32F}.
     \item[(4)] Seoul National University Bright Quasar Survey in Optical (SNUQSO). II. Discovery of 40 Bright Quasars near the Galactic plane \citep[][]{2007ApJ...664...64I}.
     \item[(5)] The LAMOST survey of Background Quasars in the Vicinity of M31 and M33 \citep[][]{2010RAA....10..612H, 2013AJ....145..159H, 2015RAA....15.1438H}.
\end{itemize}

Quasar surveys with the largest contributions to the number of spectral identification quasars, 
such as the SDSS DR16Q \citep[][]{2020ApJS..250....8L}, the early data release of Dark Energy Spectroscopic Instrument 
\citep[DESI EDR;][]{2024AJ....168...58D}, 
the 2-degree Field (2dF) QSO Redshift Survey and the associated 6-degree Field (6dF) QSO Redshift Survey
\citep[2QZ/6QZ;][]{2004MNRAS.349.1397C}, and the
LAMOST quasar survey \citep[][]{2016AJ....151...24A, 2018AJ....155..189D, 2019ApJS..240....6Y, 2023ApJS..265...25J}, 
have always avoided the area of the Galactic plane. 
Even small number of quasars located within $|b| \le 20\degr$ were included in these huge quasar catalogs, they have been
included in the Milliquas \citep[][]{2023OJAp....6E..49F}.  

\begin{deluxetable*}{llll}[htb!]
\tabletypesize{\scriptsize}
\tablecaption{Catalog format for the quasars behind the Galactic plane with $|b| \le 20\degr$ from LAMOST DR10. \label{tab:catalog}}
\tablehead{\colhead{Column} &  \colhead{Name} &  \colhead{Format} &  \colhead{Description}}
 \startdata
        1   & ObsID      &   LONG      &  Unique Spectra ID in LAMOST database \\
        2   & ObsDate  &   STRING   &  Target observation date \\
        3   & NAME      &   STRING   &  LAMOST designation hhmmss.ss+ddmmss.s (J2000) \\
        4   & RA            &   DOUBLE  &  Right ascension (R.A.) in decimal degrees (J2000) \\
        5   & DEC         &   DOUBLE  &  Declination (Decl.) in decimal degrees (J2000) \\
        6   & LMJD       &   LONG      &   Local Modified Julian Date of observation \\
        7   & PLANID    &   STRING  &   Spectroscopic plan identification \\
        8   & SPID         &   INT     &   Spectrograph identification \\
        9   & FIBERID   &   INT      &   Spectroscopic fiber identification \\
        \hline
       10 & Z$\_$VI                 &  DOUBLE  &  Redshift checked by visual inspection \\
       11  & SOURCE\_FLAG  & STRING        &  Flag of quasar candidate selection\\
       12  & M$_{i}\_$Z2              &  FLOAT  &  M$_{i} (z=2)$, absolute i-band magnitude\\ 
       13  & NSPECOBS         & INT        &  Number of spectroscopic observations for the quasar \\ 
       14  & SNR\_SPEC         &  FLOAT   &  Median S/N of the spectrum \\
       15  & EBV      & FLOAT    &  Line of sight E(B-V) from \citet{Planck_2016} dust map\\
       16  & NaD$\_$absorp     &  STRING    & Flag of `Y' (yes) or `N' (no) detection of Na D absorption lines\\ 
       17  & NOTE                   & STRING        &  Flag of `N' (new) or `K' (known) quasar \\
        \hline
        18  & gmag       &  FLOAT     &  PS1  mean PSF g magnitude \\
        19  & e\_gmag  &  FLOAT     &  g magnitude error\\
        20  & rmag        &  FLOAT    &  PS1  mean PSF r magnitude\\
        21  & e\_rmag   &   FLOAT    &  r magnitude error \\
        22  & imag        &   FLOAT    &  PS1  mean PSF i magnitude\\
        23  & e\_imag   &   FLOAT    &  i magnitude error\\
        24  & zmag       &   FLOAT    &  PS1  mean PSF z magnitude\\
        25  & e\_zmag  &   FLOAT    &  z magnitude error\\
        \hline
        26  &  PLX           &  DOUBLE  &  Parallax from Gaia DR3 \\
        27  &  e\_PLX      &   FLOAT     &  parallax error \\
        28  &  pmra          &  DOUBLE  &  PM in right ascension direction from Gaia DR3, mas/yr \\
        29  &  e\_pmra     &   FLOAT    &  pmra error  \\
        30  &  pmde         &   DOUBLE &  PM in declination direction from Gaia DR3, mas/yr \\
        31  &  e\_pmde    &   FLOAT    &  pmde error \\
        32  &  BPmag       &   FLOAT   &  Integrated BP mean magnitude from Gaia DR3\\
        33  &  e\_BPmag  &   FLOAT   &  BP magnitude error \\
        34  &  RPmag       &    FLOAT  &  Integrated RP mean magnitude from Gaia DR3\\
        35  &  e\_RPmag  &   FLOAT   &  RP magnitude error \\
        \hline
        36 & w1mpropm      &  FLOAT   & instrumental profile-fit photometry magnitude, W1 band, from AllWISE\\
        37 & e\_w1mpropm &  FLOAT   & W1 magnitude error\\
        38 & w2mpropm      &  FLOAT   & instrumental profile-fit photometry magnitude, W2 band, from AllWISE\\
        39 & e\_w2mpropm &  FLOAT   & W2 magnitude error\\
        \enddata
        \tablenotetext{}{{(This table is available in its entirety in FITS format.)}}
\end{deluxetable*}

We present a compiled catalog (Table \ref{tab:catalog}) for the quasars behind the Galactic plane with 
$|b| \le 20\degr$ in LAMOST DR10. This catalog includes all 1982 quasars for the convenience of follow-up studies. 
Among them, 1338 quasars are newly discovered, and new/known quasars are labeled in the catalog.  
This table is available on the PaperData Repository of the National Astronomical Data Center of China 
at doi:10.12149/101489\footnote{https://doi.org/10.12149/101489}.
The information for each column listed in Table~\ref{tab:catalog} is described below.

\begin{itemize}[noitemsep]
  \item[1.] Unique spectra ID in LAMOST database.
  \item[2.] Target Observation date.
  \item[3.] LAMOST DR10 target designation: $\rm Jhhmmss.ss+ddmmss.s$ (J2000). 
  \item[4-5.] Right Ascension and Declination (in decimal degrees, J2000).
  \item[6-9.] Information of the spectroscopic observation, including:
                   local modified Julian date (LMJD), spectroscopic plan identification (PLANID), spectrograph identification (SPID), 
                   and spectroscopic fiber identification (FIBERID).
                   Combining these four IDs uniquely determines a single observation of a single object. The spectral files are named in the format of
                   {spec$-$LMJD$-$PLANID\_SPID$-$FIBERID.fits}.
  \item[10.] Redshifts derived from the 1D pipeline are adopted if they concur with the visual inspections; otherwise, redshifts determined through visual inspections are used.
  \item[11.] Target selection flag.
                 This flag is a three-character string indicating how the quasar candidate is selected.
                 The first character indicates the target is selected from \citet[][]{2016ApJ...817...73H}.
                 The second character indicates the target is selected from \citet[][]{2021ApJS..254....6F}.
                 The third character indicates the target is from the two supplement origins. A single object may have more than one origins.     
  \item[12.] M$_{i}(z=2)$: absolute $i$-band magnitude. K-corrected to $z=2$ following \citet{2006AJ....131.2766R}.
  \item[13.] Number of spectroscopic observations for the quasar. Only the spectra ID with the highest S/N is listed in this catalog.  
  \item[14.] Median S/N of the spectrum.
  \item[15.] EBV: Line of sight E(B-V) from \citet[][]{Planck_2016} dust map.
  \item[16.] Na D detection flag: a flag `Y' indicates that the Na D absorption lines are detected, while `N' indicates no detections. 
  \item[17.] Note: a flag `N' indicates that the quasar is newly discovered by LAMOST, while `K' indicates that it has been listed in the literature described above.
 \item[18-25.]  PS1 mean PSF $griz_{\rm P1}$ magnitudes and errors.
 %
 \item[26-27.] Parallax and error from Gaia DR3.
 \item[28-31.] PMs and errors in right ascension and declination directions from Gaia DR3.
 \item[32-35.] Integrated BP/RP mean magnitudes and errors from Gaia DR3.
 %
 \item[36-39.] Instrumental profile-fit photometry magnitudes and errors in W1 and W2 bands from AllWISE.
 \end{itemize}

\section{Discussion} \label{sec:discussion}
 
 In this section, we will discuss the properties of this GPQ sample, including the spatial distributions, 
 the primary factors causing the non-uniform spatial distributions, magnitude and redshift distributions,
 astrometric properties and preliminary exploration of the potential applications of absorption line features 
 attributed to the ISM/IGM along the sight lines of the Milky Way.
 
\subsection{Spatial Distributions} \label{subsec:spa_distri}

\begin{figure*}[htb!]
    \centering
    \includegraphics[width=1.0\textwidth]{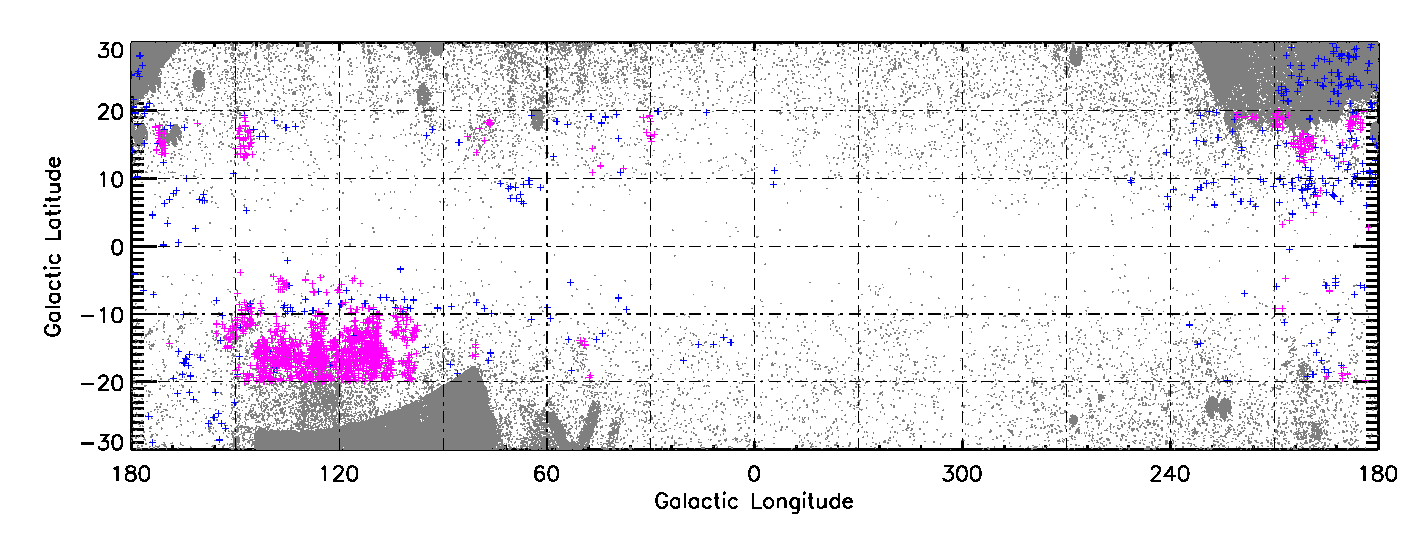}
    \caption{Spatial distributions of the 1338 LAMOST newly identified GPQs (magenta crosses) in Galactic coordinates. 
    Blue crosses represent known GPQs listed in \citet[][]{2007ApJ...664...64I}, \citet[][]{2017RAA....17...32H} and 
    \citet[][]{2022ApJS..261...32F}, while grey dots represent QSOs/AGNs or candidates from the Milliquas 
    \citep[][]{2023OJAp....6E..49F}. }
    \label{fig:GPQ_spatial}
\end{figure*}

 The spatial distributions of all known quasars behind the Galactic plane with $|b| \le 20\degr$ are shown in 
 Figure \ref{fig:GPQ_spatial}, including 1338 new GPQs identified by LAMOST (this work), known GPQs with spectroscopic identifications listed in the
 literature \citep[see][]{2007ApJ...664...64I, 2017RAA....17...32H, 2022ApJS..261...32F}, 
 as well as all known QSOs/AGNs or candidates from the Milliquas \citep[][]{2023OJAp....6E..49F}. 
 The Milliquas contains all published quasars to 30 June 2023, including quasars from
 the early data release of the DESI survey \citep[][]{2024AJ....168...58D} and the SDSS-DR18 Black Hole Mapper 
 \citep[][see the website\footnote{\url{https://data.sdss.org/sas/dr18/vac/bhm/efeds_speccomp/v1.4.3}}]{2023ApJS..267...44A}.
 There are 907,144 Type-I QSOs/AGNs, 66,026 high-confidence radio/X-ray associated quasar candidates, 2814 BL Lac objects
 and 45,816 type-II objects in Milliquas.
 As seen from Figure \ref{fig:GPQ_spatial},  the number of spectroscopically identified GPQs increased significantly with the contribution of LAMOST. 
 
In the Milliquas \citep[][]{2023OJAp....6E..49F}, the counts of Type-I QSOs/AGNs or candidates are as follows: 892 within $|b| \le 10\degr$, 3447 within $|b| \le 15\degr$, and 14,279 within $|b| \le 20\degr$. The notably larger count within $|b| \le 20\degr$, 
compared to that of $|b| \le 15\degr$ and $|b| \le 10\degr$, results from the coverage mainly provided by the 
SDSS survey in the large irregular grey area depicted in Figure \ref{fig:GPQ_spatial}, 
and the several grey circular regions (each with 8\,deg$^2$) contributed by the DESI survey.
While out of the grey area, the mean number density of known GPQs is lower than 1\,deg$^{-2}$ within $|b| \le 20\degr$, 
and even drops to $\sim$0.1\,deg$^{-2}$ within $|b| \le 10\degr$, which is extremely lower than that of the Galactic Caps,
however, the later can reach $\sim$ 100\,deg$^{-2}$.

\begin{figure*}[htb!]
    \centering
    \includegraphics[width=1.0\textwidth]{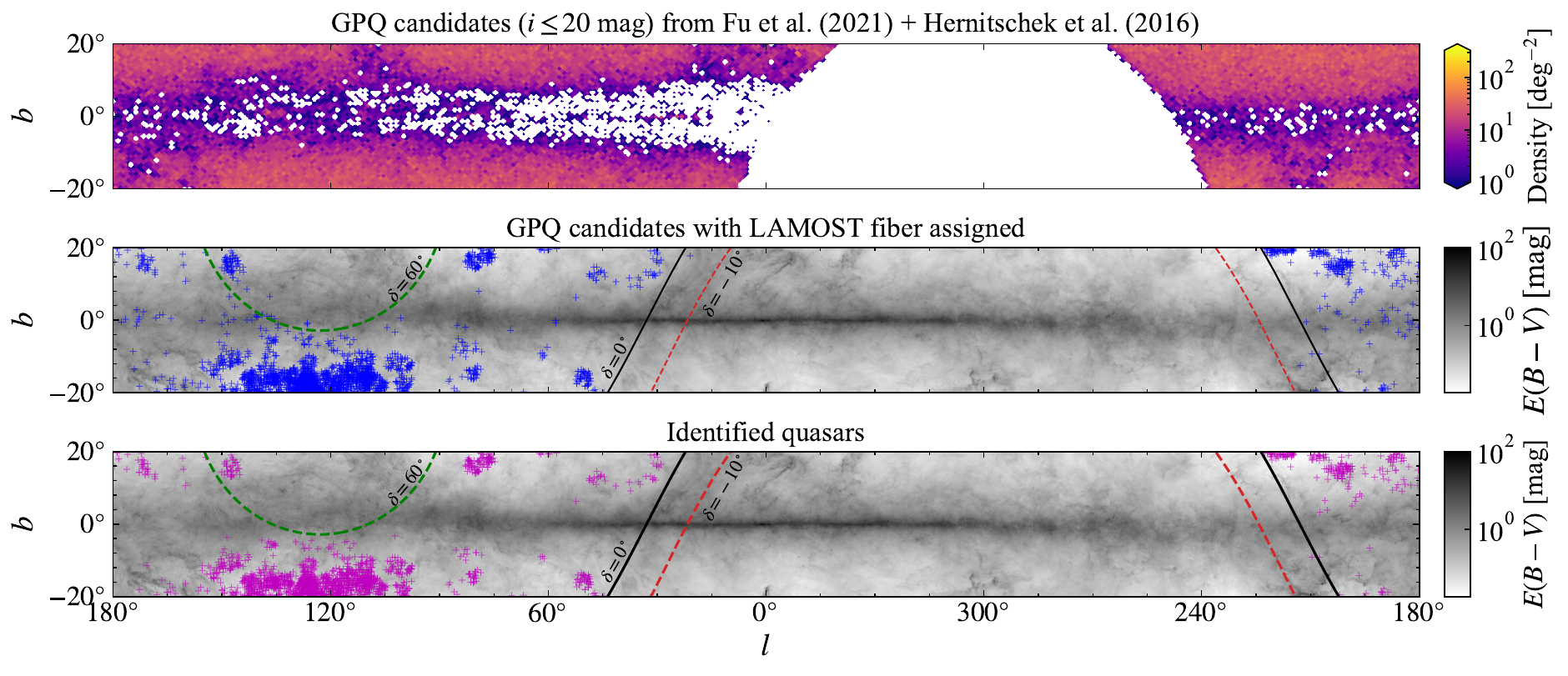}
    \caption{Sky density of the GPQ candidates from the combination list of \citet[][]{2016ApJ...817...73H}
    and \citet[][]{2021ApJS..254....6F} with Declination $\ge -30\degr$ and $i\leq 20.0$\,mag (top panel), 
    spatial distribution of the 2454 GPQ candidates with LAMOST fiber 
    assigned (middle panel), and that of the 1949 confirmed GPQs from the 2454 GPQ candidates (bottom panel). 
    The middle and bottom panels are overlaid with the \citet[][]{Planck_2016} dust map, 
    and the declination lines of $-10 \degr$, $0\degr$ and $60\degr$ are marked.}
    \label{fig:gpqc_density}
\end{figure*}

 From Figure \ref{fig:GPQ_spatial}, we can see that the spatial distribution of the LAMOST identified GPQs is highly non-uniform. 
 A significant proportion of the observation plates resulted in the identification of a limited number of quasars, 
 with over 70\% of the observation plates yielding fewer than 10 quasars. 
 Approximately 55\% of the observation plates identified either one or two quasars. 
 Around 25\% of the observation plates were relatively productive, yielding dozens of quasars.
 The three most productive observation plates yielded a substantial number of quasars, with counts of  71, 85 and 144, respectively.
 There are 1982 GPQs finally identified, with 1949 GPQs from targeted quasar candidates and 
 33 GPQs from targeted star or variable objects.

The reasons that cause the highly non-uniform spatial distributions of LAMOST GPQ are mainly the high dust extinction 
 \citep[see][]{1998ApJ...500..525S, 2016ApJ...821...78S, Planck_2016} and the crowded stellar fields 
 \citep[e.g.][]{2006AJ....131.1163S, 2009A&A...495..819R}. 
 In Figure \ref{fig:gpqc_density}, we show the sky density of the GPQ candidates combined from \citet[][]{2016ApJ...817...73H}
 and \citet[][]{2021ApJS..254....6F}, the spatial distribution of the 2454 GPQ candidates observed by LAMOST 
 until June 2022, 
 and the 1949 GPQs identified from these 2454 GPQ candidates.
The GPQ candidates are very sparse or even absent at low latitudes, which is highly relevant to the distributions of 
foreground dust extinction and stellar density. In the middle and bottom panels of Figure \ref{fig:gpqc_density}, 
the dust map from \citet[][]{Planck_2016} are overlaid to illustrate these effects.

\begin{figure*}[htb!]
    \centering
    \includegraphics[width=0.9\textwidth]{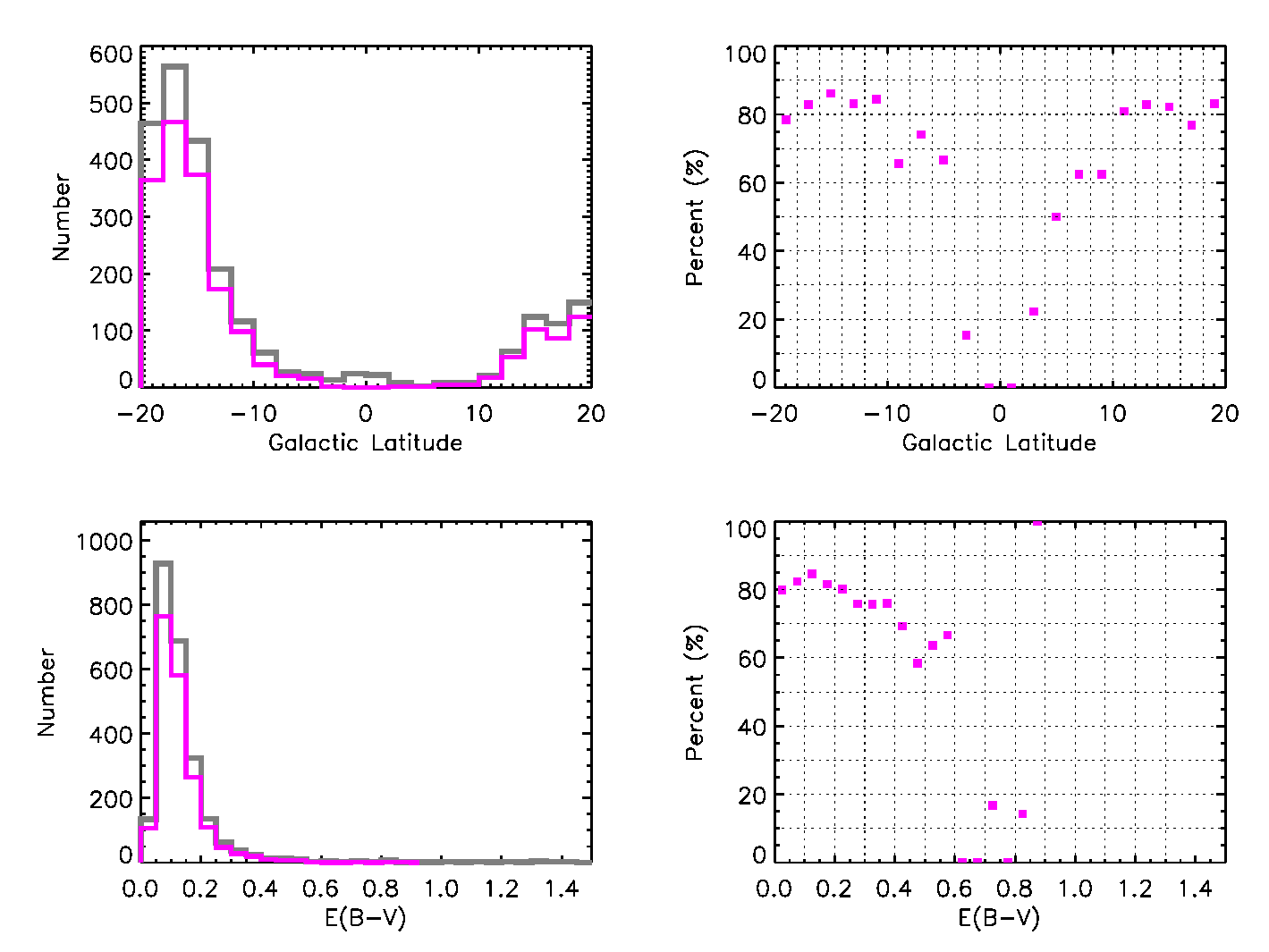}
    \caption{The distributions of the number of the LAMOST observed GPQ candidates (grey line) and the corresponding 
    confirmed GPQs (magenta line) along the Galactic Latitude (top-left) and dust extinction $E(B-V)$ (bottom-left);
    success rate of quasar identifications along the Galactic latitude (top-right) and dust extinction (bottom-right).}
    \label{fig:SR_gb_ebv}
\end{figure*}

Through simple inspection on the distributions of the number of the observed GPQ candidates and 
the confirmed GPQs along the Galactic latitude and dust extinction E(B-V), see the left panels of Figure \ref{fig:SR_gb_ebv},
we can see that the number of GPQ candidates and identified GPQs is strongly correlated with the Galactic latitude
and dust extinction. Moreover, as the Galactic latitude decreases and the dust extinction increases, the success rate
of GPQ identification decreases significantly, see the right panels of Figure \ref{fig:SR_gb_ebv}.
The success rate is relatively high, around $\sim$80\%, in the range of the Galactic latitude $10\degr<|b|<20\degr$,
corresponding to lower extinctions at the same time, with $ E(B-V) < 0.3$.
However, in the low Galactic latitude regions ($|b| < 5\degr$) where the dust extinction is usually high, with $ E(B-V) >0.6$, 
the success rate of identification is very low. It should be pointed out that 
due to the small sample size within these regions, the success rate is not well constrained. 
No GPQ was identified at regions with $E(B-V)\geq 1.0$.

In addition to the two main factors mentioned above, there are other factors involved. 
One point is that, due to the reasons of site location and design, LAMOST can only observe 
the area with $\rm{Dec} \ge -10\degr$, and at $\rm{Dec} \ge 60\degr$, the field of view is decreased to 3$\degr$.
Another factor is that the survey progress is incomplete yet, and the medium brightness observation plates with GPQ candidates
allocated need to be observed under particularly good weather conditions. 
Most LAMOST GPQs are distributed around $240\degr < l < 90\degr$. This distribution is primarily due to the suitability of 
the extended Galactic Anti-Center Area for LAMOST observations, which benefit from favorable observational conditions 
during the winter months. During this period, the Galactic Anti-Center is visible under the clearest weather conditions.
The last but not the least reason is the non-uniform performance of the 4000 fibers of LAMOST 
\citep[see][for more details]{2015RAA....15.1095L}.

\subsection{Magnitude and Redshift Distributions} \label{subsec:rz_mag }

\begin{figure*}[hbt!]
    \centering
    \includegraphics[width=1.0\textwidth]{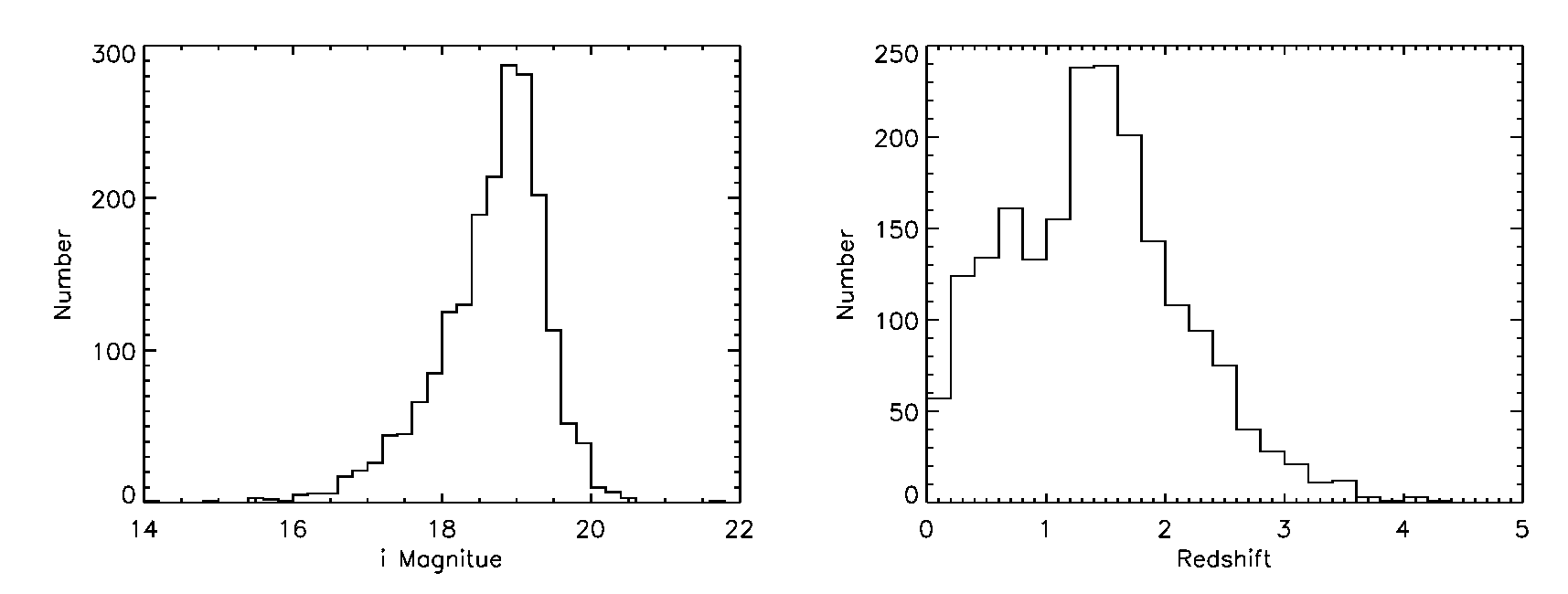}    
       \caption{PS1 $i$-band mean magnitude (left panel) and redshift (right panel) distributions of the 1982 GPQs from LAMOST DR10, 
       including 1338 newly discovered and 644 known quasars.}
    \label{fig:GPQ_imag_rz}
\end{figure*}

The left panel of Figure \ref{fig:GPQ_imag_rz} shows the PS1 mean $i$-band magnitude distributions of the 1982 GPQs from the LAMOST DR10, 
including 1338 newly discovered and 644 known quasars. The magnitude distributions have a bin size of 0.2, and the $y$-axis is the
number of quasars. The magnitude distributions have a peak at $i$-mag 19.0, and the majority of GPQs have brightness distributing around
18.0-19.5\,mag. There are 63/92/174 GPQs brighter than 17.0/17.5/18.0\,mag respectively, these bright GPQs will serve as good
candidates for absorption-line studies of the ISM/IGM along the sight lines of the Milky Way. There are 21 quasars fainter than 
20.0\,mag, but most are brighter than 20.3\,mag, lower than the magnitude limits of our quasar candidate selections. 
This is because these 21 quasars are from the supplement origins, errors from different photometric surveys and 
magnitude systems, or the differences between the mean and single-epoch magnitudes may be the causes.

The right panel of Figure \ref{fig:GPQ_imag_rz} shows the redshift distributions of this 1982 GPQ sample, from LAMOST DR10. 
The redshift has a bin in size of 0.2, and the $y$-axis is the number of quasars.
The peak of redshift distributions is around redshift $\sim$1.5, and most quasars have redshifts between 0.3 and 2.5. There are 52/11/4 quasars
with redshift higher than 3.0/3.5/4.0. The object with the highest redshift in this sample is quasar J185726.26+463030.1, which
is a known quasar at redshift $z = 4.211$ recorded in the Milliquas \citep[][]{2023OJAp....6E..49F},
while the highest redshift quasar newly identified is J235025.26+455121.2, at $z=4.163$.
However the trough near redshift $\sim$1.0 still remains, 
induced by the inefficient identification near this redshift range because the Mg II emission line falls into the overlap range of 
the blue and red spectroscopic channels, as discussed in section \ref{sec:Iden_cat}.

\subsection{Astrometry} \label{subsec:astrometry }

\begin{figure*}[htb!]
    \centering
    \includegraphics[width=1.0\textwidth]{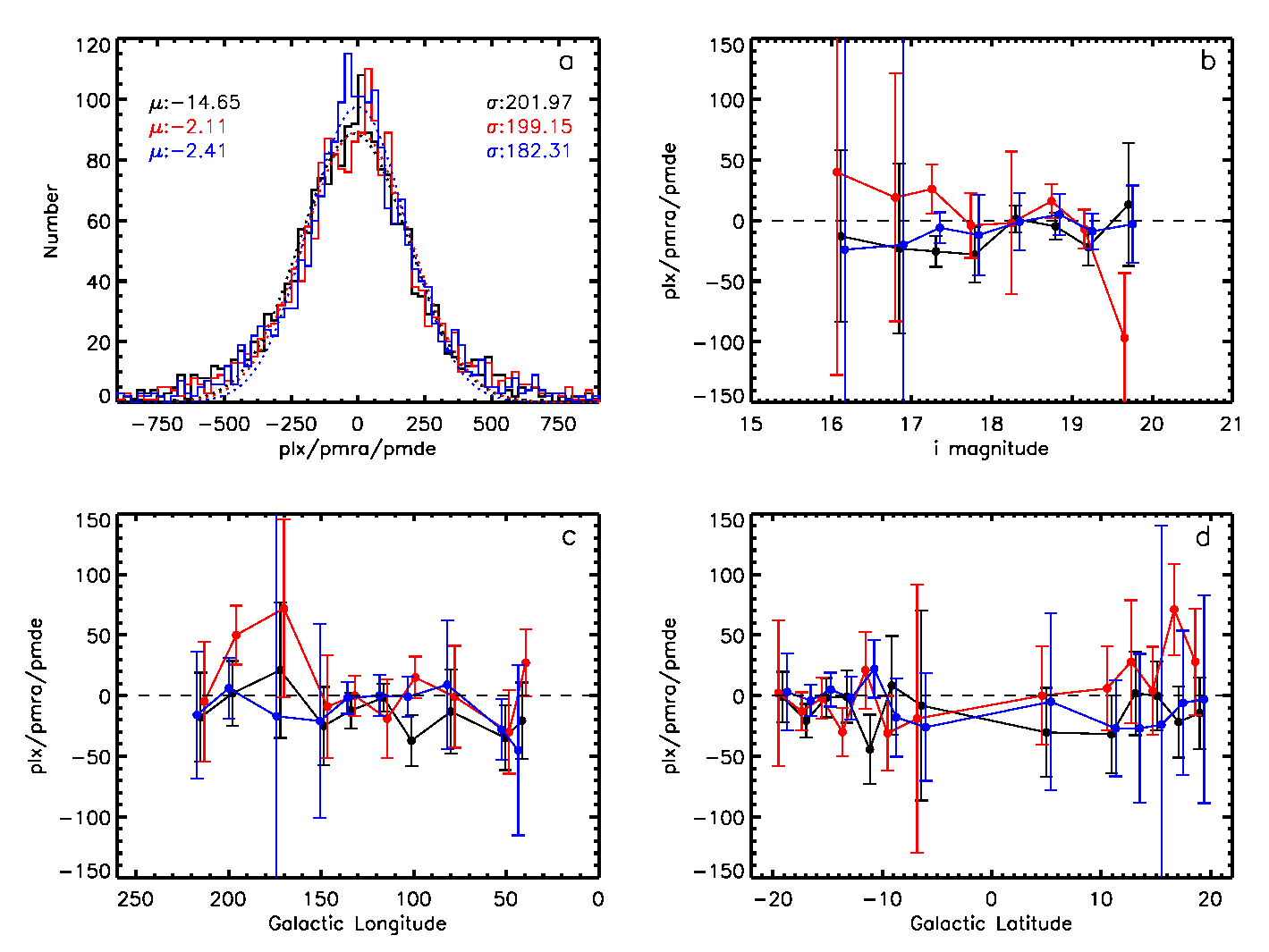}
    \caption{Panel a: Distributions of parallax (black solid line ), PM of right ascension (red solid line) and declination (blue solid line)
    from Gaia DR3, and their best-fitting Gaussian models (dotted lines) for the 1982 GPQs from LAMOST DR10. 
    The median and standard deviation values are labeled with corresponding colors.
    Panel b/c/d: Distributions of parallax (black), PM of right ascension (red) and declination (blue) 
    along the i-magnitude, the Galactic longitude and the Galactic latitude. 
    In these three panels, the bin is same for the parallax, PM of right ascension and declination, the data points are offset for clarity
    in the plot. The number of GPQs in each bin is no less than 20. The unit is $\mu$as for the parallax, and $\mu$as\,yr$^{-1}$ 
    for the PM in right ascension and declination. }
    \label{fig:GPQ_astrometry}
\end{figure*}

Quasars provide the most direct way to estimate the precision and accuracy of astrometric measurements.
This GPQ sample is vital for calibrating the parallax and PM systematics for existing Gaia releases and future releases.
As an examination, we show the distributions of the parallax, PMs of right ascension and declination from Gaia DR3 
\citep[][]{Gaia_DR3_2022} in the panel a of Figure \ref{fig:GPQ_astrometry}.
The median offsets are $-14.65$\,$\mu$as for the parallax, $-2.11$ and $-2.41$\,$\mu$as\,yr$^{-1}$ for the PMs of 
right ascension and declination. All three parameters have relatively large scatters. 
The median zero point offset of Gaia parallax is close to the value of $-12.7$\,$\mu$as obtained by \citet{2024A&A...691A..81D} using a refined compilation of quasar candidates at $|b|\leq20\degr$. Both zero-point offsets of parallax from this work and \citet{2024A&A...691A..81D} are slightly smaller than the all-sky median parallax bias of $-17$\,$\mu$as reported by \citet{2021A&A...649A...4L}. As shown by \citet{2021A&A...649A...4L}, the parallax bias is dependent on the magnitude, 
color, and position of the source, with systematic variations at a level of $\sim$10\,$\mu$as. Still, the variations are most 
significant along the Galactic plane -- the zone of avoidance for quasar survey (see their Section 4.1).

We also examined the distributions of the parallax, PMs of right ascension and declination as a function of i magnitude,
Galactic longitude and Galactic latitude separately, see the panels b, c, and d in Figure \ref{fig:GPQ_astrometry}. 
The number of GPQs is limited, but the number of each bin is no less than 20. 
The number of GPQs in magnitude range  i-mag $< 17$, near the Galactic longitude 170$\degr$ and in Galactic latitude 
$-8\degr < b < 10\degr$ is minimal accompanied by larger error bars.
However, we still can see significant patterns, e.g., the non-zero offsets, the rise and fall, in the distributions of the three 
astrometric parameters along the magnitude and position.
This sample, together with more QSOs behind the Galactic plane to be discovered in the future, 
could validate and improve the astrometry of Gaia.

\subsection{Absorption Lines} \label{subsec:absorp }

The absorption line characteristics of the GPQ spectra carry the physical information of the ISM/IGM along the lines of sight.
In the LAMOST wavelength coverage, the Ca~\textsc{ii} and Na~\textsc{i} lines are the most prominent, corresponding to the 
H (3934.78\AA), K (3969.59\AA) and D (D$_2$: 5891.58\AA, D$_1$: 5897.56\AA) lines discovered by Fraunhofer 
two hundred years ago \citep[][]{fraunhofer1817bestimmung}.
Since LAMOST spectra have poor throughput bluewards of 4000\AA, where the H, and K lines are located,
only preliminary results of Na D absorption lines are discussed here.   
In addition, the Na D lines fall in the overlap region of the blue and red channels (5700-5900\AA), where 
the throughputs are relatively lower and artifacts may be induced during the blue and red channel spectra splicing. 
Another factor that can affect the analysis of Na D absorption lines in the LAMOST spectra, is that there are strong Na D emission lines
in the terrestrial nightglow produced from the reaction between Na and O$^3$ \citep[][]{2012JASTP..74..181P}.
Therefore, there are certain difficulties in conducting detailed scientific research on Na D absorption characteristics in LAMOST spectra.

\begin{figure*}[bht!]
    \centering
    \includegraphics[width=1.0\textwidth]{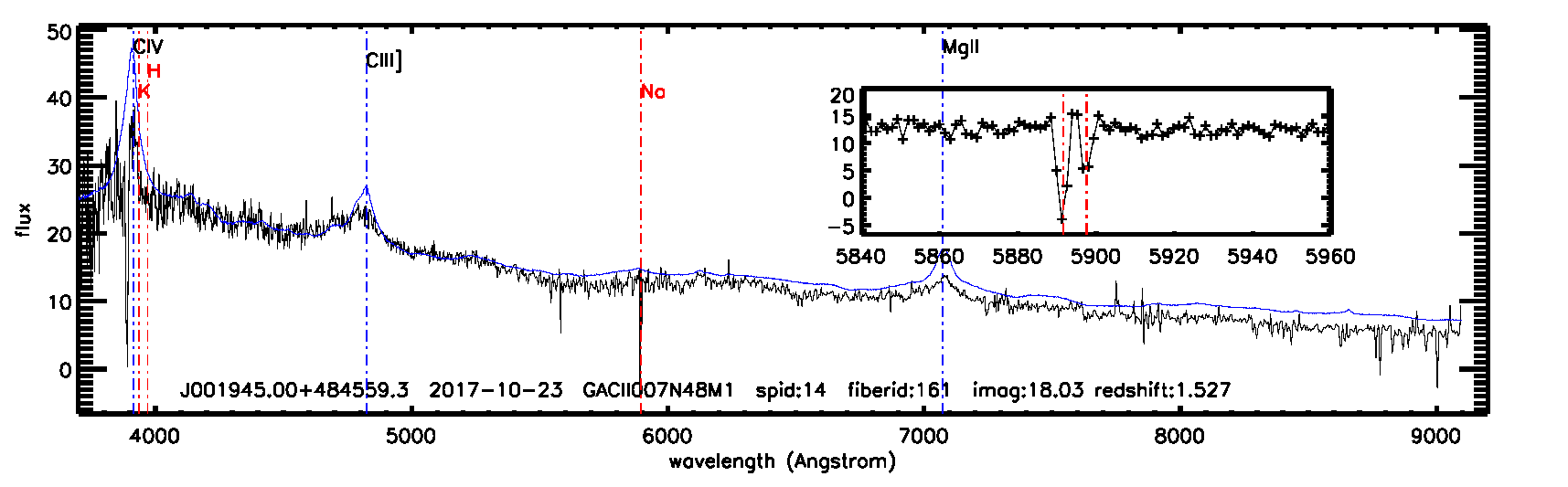}
    \caption{LAMOST Spectrum of quasar J001945.00+484559.3 (black) at redshift $z=1.527$ with composite quasar spectrum overlaid
    \citep[blue solid line, ][]{2001AJ....122..549V}, and identified emission lines labeled (blue dash-dotted line).
   The red dash-dotted lines indicate the laboratory wavelength positions of the Ca H\&K and Na D doublets.
    The zoom-in plot shows the spectrum of the wavelength range 5840-5960\AA.}
    \label{fig:spec_absorp}
\end{figure*}

Upon visual examination of the 1982 LAMOST GPQ spectra, the Na D absorption line characteristics were detected in 
approximately 50\% of the GPQ spectra. 
This ratio is much higher compared to that of high Galactic latitude targets, and this result is reasonable. 
The strength of Na D absorption line is generally expected to correlate with the amount of dust along the line of sight.
\citet[][]{2012MNRAS.426.1465P} presented a detailed study of the strong correlation between the dust extinction 
and the equivalent width (EW) of the Na D absorption doublet with more than 100 Keck high-resolution spectra and nearly 
a million SDSS low-resolution spectra.
\citet[][]{2015MNRAS.452..511M} mapped out Ca H\&K and Na D absorption lines induced by the ISM and the circumgalactic medium 
of the Milky Way with the stacked SDSS extragalactic spectra in Galactic latitude bins, and their results show both the Ca H\&K and Na D absorption increase from high Galactic latitude towards low Galactic latitude.
Figure \ref{fig:spec_absorp} shows one example spectrum of LAMOST GPQ, J001945.00+484559.3, at redshift $z=1.527$ with clearly
Na D absorption lines detected.

\begin{figure}[htb!]
    \centering
    \includegraphics[width=0.5\textwidth]{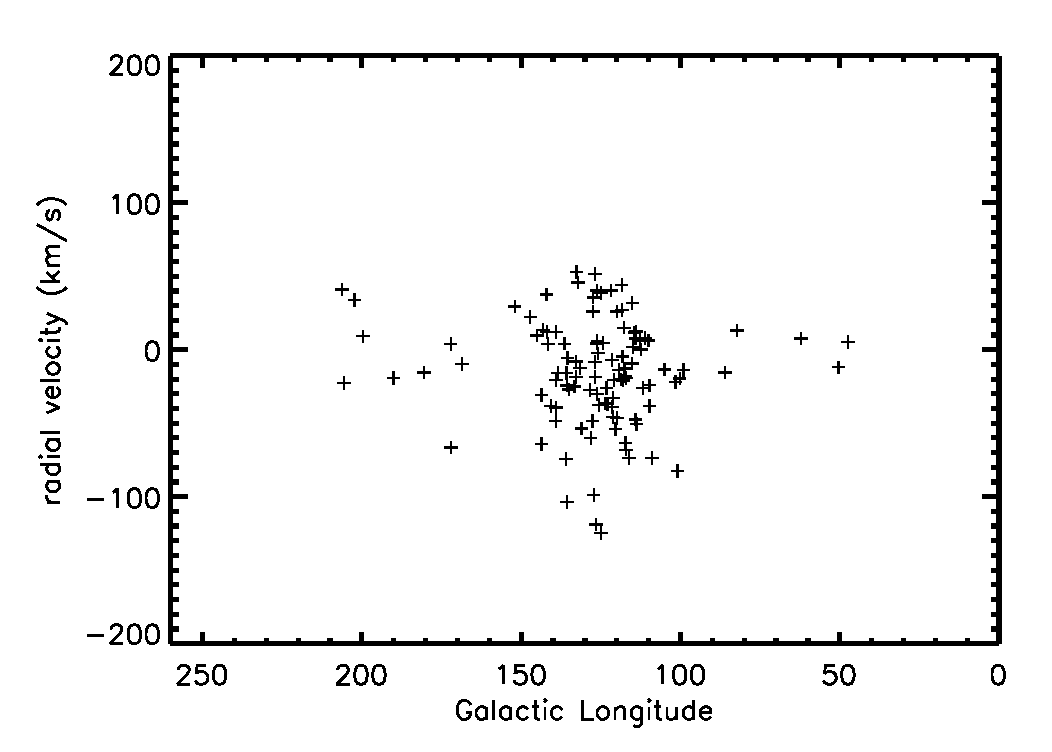}
    \caption{Radial velocities along the Galactic longitude measured from D$_2$ lines for about one hundred LAMOST GPQs with relatively better Na D absorption
    line detections.}
    \label{fig:absorp_vel}
\end{figure}

Because most of LAMOST GPQs have relatively low S/Ns in this wavelength coverage, and a portion of spectra is 
contaminated by the Na D emission in the terrestrial nightglow, for the vast majority of spectra we could not get 
a double Gaussian fit to converge and give a useful measurement. Only distributions of radial velocities of Na D absorption 
lines along the Galactic longitude for approximately one hundred GPQ spectra are shown here to demonstrate the potential applications in the future.
For clarity, only radial velocities obtained from D$_2$ lines are plotted in Figure \ref{fig:absorp_vel}.
In the future, the spectra of higher quality and resolution will help constrain the properties of the ISM/IGM of the Milky Way precisely
along the lines of sight of GPQs.

\section{Summary} \label{sec:summary}

In this work, we report the discovery of 1982 quasars behind the Galactic plane region with $|b| \le 20\degr$, 
utilizing data from the LAMOST DR10. Among these, 1338 quasars are newly identified. 
The number of spectroscopically identified GPQs increased significantly with the contribution of LAMOST.
The selection of quasar candidates was based on their variability and color characteristics, derived from the optical and infrared 
photometric data through the PS1 and WISE surveys.
The spatial distribution of this GPQ sample exhibits significant non-uniformity, attributed to the candidate selections, 
observational planning progress and instrumental effects.
Additionally, the magnitude and redshift distributions show that this GPQ sample is a relatively bright sample with intermediate redshifts. 
Furthermore, we investigate the astrometric properties, and explore the implications of this sample for potential applications 
for absorption line analysis, which could provide insights into the ISM/IGM of the Milky Way along the lines of sight. 
We expect to identify approximately 5000 more GPQs with $|b|<20\degr$ from the ongoing LAMOST phase III survey.

\begin{acknowledgments}

We would like to thank the reviewer for his/her valuable comments and suggestions, which have helped to improve 
the quality of this paper. 
This work is supported by the National Natural Science Foundation of China 12133001.
The Guoshoujing Telescope (the Large Sky Area Multi-object Fiber Spectroscopic Telescope LAMOST) is a National Major 
Scientific Project built by the Chinese Academy of Sciences. Funding for the project has been provided by the National 
Development and Reform Commission. LAMOST is operated and managed by the National Astronomical Observatories, 
Chinese Academy of Sciences.
The Pan-STARRS1 Surveys (PS1) and the PS1 public science archive have been made possible through contributions by the Institute for Astronomy, the University of Hawaii, the Pan-STARRS Project Office, the Max-Planck Society and its participating institutes, the Max Planck Institute for Astronomy, Heidelberg and the Max Planck Institute for Extraterrestrial Physics, Garching, The Johns Hopkins University, Durham University, the University of Edinburgh, the Queen's University Belfast, the Harvard-Smithsonian Center for Astrophysics, the Las Cumbres Observatory Global Telescope Network Incorporated, the National Central University of Taiwan, the Space Telescope Science Institute, the National Aeronautics and Space Administration under Grant No. NNX08AR22G issued through the Planetary Science Division of the NASA Science Mission Directorate, the National Science Foundation Grant No. AST-1238877, the University of Maryland, Eotvos Lorand University (ELTE), the Los Alamos National Laboratory, and the Gordon and Betty Moore Foundation.
This publication makes use of data products from the Wide-field Infrared Survey Explorer, which is a joint project of the University of California, Los Angeles, and the Jet Propulsion Laboratory/California Institute of Technology, funded by the National Aeronautics and Space Administration.
This work presents results from the European Space Agency (ESA) space mission Gaia. Gaia data are being processed by the Gaia Data Processing and Analysis Consortium (DPAC). Funding for the DPAC is provided by national institutions, in particular the institutions participating in the Gaia MultiLateral Agreement (MLA). The Gaia mission website is https://www.cosmos.esa.int/gaia. The Gaia archive website is https://archives.esac.esa.int/gaia.

\end{acknowledgments}


\bibliography{qso_Galactic_plane}{}

\begin{thebibliography}{}
\expandafter\ifx\csname natexlab\endcsname\relax\def\natexlab#1{#1}\fi
\providecommand{\url}[1]{\href{#1}{#1}}
\providecommand{\dodoi}[1]{doi:~\href{http://doi.org/#1}{\nolinkurl{#1}}}
\providecommand{\doeprint}[1]{\href{http://ascl.net/#1}{\nolinkurl{http://ascl.net/#1}}}
\providecommand{\doarXiv}[1]{\href{https://arxiv.org/abs/#1}{\nolinkurl{https://arxiv.org/abs/#1}}}

\bibitem[{{Abuter} {et~al.}(2024){Abuter}, {Allouche}, {Amorim}, {Bailet},
  {Berdeu}, {Berger}, {Berio}, {Bigioli}, {Boebion}, {Bolzer}, {Bonnet},
  {Bourdarot}, {Bourget}, {Brandner}, {Cao}, {Conzelmann}, {Comin},
  {Cl{\'e}net}, {Courtney-Barrer}, {Davies}, {Defr{\`e}re}, {Delboulb{\'e}},
  {Delplancke-Str{\"o}bele}, {Dembet}, {Dexter}, {de Zeeuw}, {Drescher},
  {Eckart}, {{\'E}douard}, {Eisenhauer}, {Fabricius}, {Feuchtgruber}, {Finger},
  {F{\"o}rster Schreiber}, {Garcia}, {Garcia Lopez}, {Gao}, {Gendron},
  {Genzel}, {Gil}, {Gillessen}, {Gomes}, {Gont{\'e}}, {Gouvret}, {Guajardo},
  {Guieu}, {Hackenberg}, {Haddad}, {Hartl}, {Haubois}, {Hau{\ss}mann},
  {Hei{\ss}el}, {Henning}, {Hippler}, {H{\"o}nig}, {Horrobin}, {Hubin},
  {Jacqmart}, {Jocou}, {Kaufer}, {Kervella}, {Kolb}, {Korhonen}, {Lacour},
  {Lagarde}, {Lai}, {Lapeyr{\`e}re}, {Laugier}, {Le Bouquin}, {Leftley},
  {L{\'e}na}, {Lewis}, {Liu}, {Lopez}, {Lutz}, {Magnard}, {Mang}, {Marcotto},
  {Maurel}, {M{\'e}rand}, {Millour}, {More}, {Netzer}, {Nowacki}, {Nowak},
  {Oberti}, {Ott}, {Pallanca}, {Paumard}, {Perraut}, {Perrin}, {Petrov},
  {Pfuhl}, {Pourr{\'e}}, {Rabien}, {Rau}, {Riquelme}, {Robbe-Dubois}, {Rochat},
  {Salman}, {Sanchez-Bermudez}, {Santos}, {Scheithauer}, {Sch{\"o}ller},
  {Schubert}, {Schuhler}, {Shangguan}, {Shchekaturov}, {Shimizu}, {Sevin},
  {Soulez}, {Spang}, {Stadler}, {Sternberg}, {Straubmeier}, {Sturm}, {Sykes},
  {Tacconi}, {Tristram}, {Vincent}, {von Fellenberg}, {Uysal}, {Widmann},
  {Wieprecht}, {Wiezorrek}, {Woillez}, \& {Zins}}]{2024Natur.627..281A}
{Abuter}, R., {Allouche}, F., {Amorim}, A., {et~al.} 2024, \nat, 627, 281,
  \dodoi{10.1038/s41586-024-07053-4}

\bibitem[{{Ahn} {et~al.}(2012){Ahn}, {Alexandroff}, {Allende Prieto},
  {Anderson}, {Anderton}, {Andrews}, {Aubourg}, {Bailey}, {Balbinot}, {Barnes},
  {Bautista}, {Beers}, {Beifiori}, {Berlind}, {Bhardwaj}, {Bizyaev}, {Blake},
  {Blanton}, {Blomqvist}, {Bochanski}, {Bolton}, {Borde}, {Bovy}, {Brandt},
  {Brinkmann}, {Brown}, {Brownstein}, {Bundy}, {Busca}, {Carithers}, {Carnero},
  {Carr}, {Casetti-Dinescu}, {Chen}, {Chiappini}, {Comparat}, {Connolly},
  {Crepp}, {Cristiani}, {Croft}, {Cuesta}, {da Costa}, {Davenport}, {Dawson},
  {de Putter}, {De Lee}, {Delubac}, {Dhital}, {Ealet}, {Ebelke}, {Edmondson},
  {Eisenstein}, {Escoffier}, {Esposito}, {Evans}, {Fan}, {Femen{\'\i}a
  Castell{\'a}}, {Fern{\'a}ndez Alvar}, {Ferreira}, {Filiz Ak}, {Finley},
  {Fleming}, {Font-Ribera}, {Frinchaboy}, {Garc{\'\i}a-Hern{\'a}ndez},
  {Garc{\'\i}a P{\'e}rez}, {Ge}, {G{\'e}nova-Santos}, {Gillespie}, {Girardi},
  {Gonz{\'a}lez Hern{\'a}ndez}, {Grebel}, {Gunn}, {Guo}, {Haggard}, {Hamilton},
  {Harris}, {Hawley}, {Hearty}, {Ho}, {Hogg}, {Holtzman}, {Honscheid},
  {Huehnerhoff}, {Ivans}, {Ivezi{\'c}}, {Jacobson}, {Jiang}, {Johansson},
  {Johnson}, {Kauffmann}, {Kirkby}, {Kirkpatrick}, {Klaene}, {Knapp}, {Kneib},
  {Le Goff}, {Leauthaud}, {Lee}, {Lee}, {Long}, {Loomis}, {Lucatello},
  {Lundgren}, {Lupton}, {Ma}, {Ma}, {MacDonald}, {Mack}, {Mahadevan}, {Maia},
  {Majewski}, {Makler}, {Malanushenko}, {Malanushenko}, {Manchado},
  {Mandelbaum}, {Manera}, {Maraston}, {Margala}, {Martell}, {McBride},
  {McGreer}, {McMahon}, {M{\'e}nard}, {Meszaros}, {Miralda-Escud{\'e}},
  {Montero-Dorta}, {Montesano}, {Morrison}, {Muna}, {Munn}, {Murayama},
  {Myers}, {Neto}, {Nguyen}, {Nichol}, {Nidever}, {Noterdaeme}, {Nuza},
  {Ogando}, {Olmstead}, {Oravetz}, {Owen}, {Padmanabhan},
  {Palanque-Delabrouille}, {Pan}, {Parejko}, {Parihar}, {P{\^a}ris},
  {Pattarakijwanich}, {Pepper}, {Percival}, {P{\'e}rez-Fournon},
  {P{\'e}rez-R{\`a}fols}, {Petitjean}, {Pforr}, {Pieri}, {Pinsonneault}, {Porto
  de Mello}, {Prada}, {Price-Whelan}, {Raddick}, {Rebolo}, {Rich}, {Richards},
  {Robin}, {Rocha-Pinto}, {Rockosi}, {Roe}, {Ross}, {Ross}, {Rossi},
  {Rubi{\~n}o-Martin}, {Samushia}, {Sanchez Almeida}, {S{\'a}nchez},
  {Santiago}, {Sayres}, {Schlegel}, {Schlesinger}, {Schmidt}, {Schneider},
  {Schultheis}, {Schwope}, {Sc{\'o}ccola}, {Seljak}, {Sheldon}, {Shen}, {Shu},
  {Simmerer}, {Simmons}, {Skibba}, {Skrutskie}, {Slosar}, {Sobreira}, {Sobeck},
  {Stassun}, {Steele}, {Steinmetz}, {Strauss}, {Streblyanska}, {Suzuki},
  {Swanson}, {Tal}, {Thakar}, {Thomas}, {Thompson}, {Tinker}, {Tojeiro},
  {Tremonti}, {Vargas Maga{\~n}a}, {Verde}, {Viel}, {Vikas}, {Vogt}, {Wake},
  {Wang}, {Weaver}, {Weinberg}, {Weiner}, {West}, {White}, {Wilson},
  {Wisniewski}, {Wood-Vasey}, {Yanny}, {Y{\`e}che}, {York}, {Zamora},
  {Zasowski}, {Zehavi}, {Zhao}, {Zheng}, {Zhu}, \&
  {Zinn}}]{2012ApJS..203...21A}
{Ahn}, C.~P., {Alexandroff}, R., {Allende Prieto}, C., {et~al.} 2012, \apjs,
  203, 21, \dodoi{10.1088/0067-0049/203/2/21}

\bibitem[{{Ai} {et~al.}(2016){Ai}, {Wu}, {Yang}, {Yang}, {Wang}, {Guo}, {Zuo},
  {Dong}, {Zhang}, {Yuan}, {Song}, {Wang}, {Dong}, {Yang}, {-Wu}, {Shen},
  {Shi}, {He}, {Lei}, {Li}, {Luo}, {Zhao}, \& {Zhang}}]{2016AJ....151...24A}
{Ai}, Y.~L., {Wu}, X.-B., {Yang}, J., {et~al.} 2016, \aj, 151, 24,
  \dodoi{10.3847/0004-6256/151/2/24}

\bibitem[{{Aihara} {et~al.}(2011){Aihara}, {Allende Prieto}, {An}, {Anderson},
  {Aubourg}, {Balbinot}, {Beers}, {Berlind}, {Bickerton}, {Bizyaev}, {Blanton},
  {Bochanski}, {Bolton}, {Bovy}, {Brandt}, {Brinkmann}, {Brown}, {Brownstein},
  {Busca}, {Campbell}, {Carr}, {Chen}, {Chiappini}, {Comparat}, {Connolly},
  {Cortes}, {Croft}, {Cuesta}, {da Costa}, {Davenport}, {Dawson}, {Dhital},
  {Ealet}, {Ebelke}, {Edmondson}, {Eisenstein}, {Escoffier}, {Esposito},
  {Evans}, {Fan}, {Femen{\'\i}a Castell{\'a}}, {Font-Ribera}, {Frinchaboy},
  {Ge}, {Gillespie}, {Gilmore}, {Gonz{\'a}lez Hern{\'a}ndez}, {Gott}, {Gould},
  {Grebel}, {Gunn}, {Hamilton}, {Harding}, {Harris}, {Hawley}, {Hearty}, {Ho},
  {Hogg}, {Holtzman}, {Honscheid}, {Inada}, {Ivans}, {Jiang}, {Johnson},
  {Jordan}, {Jordan}, {Kazin}, {Kirkby}, {Klaene}, {Knapp}, {Kneib},
  {Kochanek}, {Koesterke}, {Kollmeier}, {Kron}, {Lampeitl}, {Lang}, {Le Goff},
  {Lee}, {Lin}, {Long}, {Loomis}, {Lucatello}, {Lundgren}, {Lupton}, {Ma},
  {MacDonald}, {Mahadevan}, {Maia}, {Makler}, {Malanushenko}, {Malanushenko},
  {Mandelbaum}, {Maraston}, {Margala}, {Masters}, {McBride}, {McGehee},
  {McGreer}, {M{\'e}nard}, {Miralda-Escud{\'e}}, {Morrison}, {Mullally},
  {Muna}, {Munn}, {Murayama}, {Myers}, {Naugle}, {Neto}, {Nguyen}, {Nichol},
  {O'Connell}, {Ogando}, {Olmstead}, {Oravetz}, {Padmanabhan},
  {Palanque-Delabrouille}, {Pan}, {Pandey}, {P{\^a}ris}, {Percival},
  {Petitjean}, {Pfaffenberger}, {Pforr}, {Phleps}, {Pichon}, {Pieri}, {Prada},
  {Price-Whelan}, {Raddick}, {Ramos}, {Reyl{\'e}}, {Rich}, {Richards}, {Rix},
  {Robin}, {Rocha-Pinto}, {Rockosi}, {Roe}, {Rollinde}, {Ross}, {Ross},
  {Rossetto}, {S{\'a}nchez}, {Sayres}, {Schlegel}, {Schlesinger}, {Schmidt},
  {Schneider}, {Sheldon}, {Shu}, {Simmerer}, {Simmons}, {Sivarani}, {Snedden},
  {Sobeck}, {Steinmetz}, {Strauss}, {Szalay}, {Tanaka}, {Thakar}, {Thomas},
  {Tinker}, {Tofflemire}, {Tojeiro}, {Tremonti}, {Vandenberg}, {Vargas
  Maga{\~n}a}, {Verde}, {Vogt}, {Wake}, {Wang}, {Weaver}, {Weinberg}, {White},
  {White}, {Yanny}, {Yasuda}, {Yeche}, \& {Zehavi}}]{2011ApJS..193...29A}
{Aihara}, H., {Allende Prieto}, C., {An}, D., {et~al.} 2011, \apjs, 193, 29,
  \dodoi{10.1088/0067-0049/193/2/29}

\bibitem[{{Almeida} {et~al.}(2023){Almeida}, {Anderson},
  {Argudo-Fern{\'a}ndez}, {Badenes}, {Barger}, {Barrera-Ballesteros}, {Bender},
  {Benitez}, {Besser}, {Bird}, {Bizyaev}, {Blanton}, {Bochanski}, {Bovy},
  {Brandt}, {Brownstein}, {Buchner}, {Bulbul}, {Burchett}, {Cano D{\'\i}az},
  {Carlberg}, {Casey}, {Chandra}, {Cherinka}, {Chiappini}, {Coker}, {Comparat},
  \& et~al.}]{2023ApJS..267...44A}
{Almeida}, A., {Anderson}, S.~F., {Argudo-Fern{\'a}ndez}, M., {et~al.} 2023,
  \apjs, 267, 44, \dodoi{10.3847/1538-4365/acda98}

\bibitem[{{Arenou} {et~al.}(2018){Arenou}, {Luri}, {Babusiaux}, {Fabricius},
  {Helmi}, {Muraveva}, {Robin}, {Spoto}, {Vallenari}, {Antoja},
  {Cantat-Gaudin}, {Jordi}, {Leclerc}, {Reyl{\'e}}, {Romero-G{\'o}mez}, {Shih},
  {Soria}, {Barache}, {Bossini}, {Bragaglia}, {Breddels}, {Fabrizio},
  {Lambert}, {Marrese}, {Massari}, {Moitinho}, {Robichon}, {Ruiz-Dern},
  {Sordo}, {Veljanoski}, {Eyer}, {Jasniewicz}, {Pancino}, {Soubiran}, {Spagna},
  {Tanga}, {Turon}, \& {Zurbach}}]{2018A&A...616A..17A}
{Arenou}, F., {Luri}, X., {Babusiaux}, C., {et~al.} 2018, \aap, 616, A17,
  \dodoi{10.1051/0004-6361/201833234}

\bibitem[{{Babusiaux} {et~al.}(2023){Babusiaux}, {Fabricius}, {Khanna},
  {Muraveva}, {Reyl{\'e}}, {Spoto}, {Vallenari}, {Luri}, {Arenou},
  {{\'A}lvarez}, {Anders}, {Antoja}, {Balbinot}, {Barache}, {Bauchet},
  {Bossini}, {Busonero}, {Cantat-Gaudin}, {Carrasco}, {Dafonte}, {Diakit{\'e}},
  {Figueras}, {Garcia-Gutierrez}, {Garofalo}, {Helmi}, {Jim{\'e}nez-Arranz},
  {Jordi}, {Kervella}, {Kostrzewa-Rutkowska}, {Leclerc}, {Licata}, {Manteiga},
  {Masip}, {Mongui{\'o}}, {Ramos}, {Robichon}, {Robin}, {Romero-G{\'o}mez},
  {S{\'a}ez}, {Santove{\~n}a}, {Spina}, {Torralba Elipe}, \&
  {Weiler}}]{2023A&A...674A..32B}
{Babusiaux}, C., {Fabricius}, C., {Khanna}, S., {et~al.} 2023, \aap, 674, A32,
  \dodoi{10.1051/0004-6361/202243790}

\bibitem[{{Ben Bekhti} {et~al.}(2008){Ben Bekhti}, {Richter}, {Westmeier}, \&
  {Murphy}}]{2008A&A...487..583B}
{Ben Bekhti}, N., {Richter}, P., {Westmeier}, T., \& {Murphy}, M.~T. 2008,
  \aap, 487, 583, \dodoi{10.1051/0004-6361:20079067}

\bibitem[{{Ben Bekhti} {et~al.}(2012){Ben Bekhti}, {Winkel}, {Richter}, {Kerp},
  {Klein}, \& {Murphy}}]{2012A&A...542A.110B}
{Ben Bekhti}, N., {Winkel}, B., {Richter}, P., {et~al.} 2012, \aap, 542, A110,
  \dodoi{10.1051/0004-6361/201118673}

\bibitem[{{Blomqvist} {et~al.}(2019){Blomqvist}, {du Mas des Bourboux},
  {Busca}, {de Sainte Agathe}, {Rich}, {Balland}, {Bautista}, {Dawson},
  {Font-Ribera}, {Guy}, {Le Goff}, {Palanque-Delabrouille}, {Percival},
  {P{\'e}rez-R{\`a}fols}, {Pieri}, {Schneider}, {Slosar}, \&
  {Y{\`e}che}}]{2019A&A...629A..86B}
{Blomqvist}, M., {du Mas des Bourboux}, H., {Busca}, N.~G., {et~al.} 2019,
  \aap, 629, A86, \dodoi{10.1051/0004-6361/201935641}

\bibitem[{{Bogd{\'a}n} {et~al.}(2024){Bogd{\'a}n}, {Goulding}, {Natarajan},
  {Kov{\'a}cs}, {Tremblay}, {Chadayammuri}, {Volonteri}, {Kraft}, {Forman},
  {Jones}, {Churazov}, \& {Zhuravleva}}]{2024NatAs...8..126B}
{Bogd{\'a}n}, {\'A}., {Goulding}, A.~D., {Natarajan}, P., {et~al.} 2024, Nature
  Astronomy, 8, 126, \dodoi{10.1038/s41550-023-02111-9}

\bibitem[{{Brown}(2021)}]{2021ARA&A..59...59B}
{Brown}, A. G.~A. 2021, \araa, 59, 59,
  \dodoi{10.1146/annurev-astro-112320-035628}

\bibitem[{{Chambers} {et~al.}(2016){Chambers}, {Magnier}, {Metcalfe},
  {Flewelling}, {Huber}, {Waters}, {Denneau}, {Draper}, {Farrow}, {Finkbeiner},
  {Holmberg}, {Koppenhoefer}, {Price}, {Rest}, {Saglia}, {Schlafly}, {Smartt},
  {Sweeney}, {Wainscoat}, {Burgett}, {Chastel}, {Grav}, {Heasley}, {Hodapp},
  {Jedicke}, {Kaiser}, {Kudritzki}, {Luppino}, {Lupton}, {Monet}, {Morgan},
  {Onaka}, {Shiao}, {Stubbs}, {Tonry}, {White}, {Ba{\~n}ados}, {Bell},
  {Bender}, {Bernard}, {Boegner}, {Boffi}, {Botticella}, {Calamida},
  {Casertano}, {Chen}, {Chen}, {Cole}, {Deacon}, {Frenk}, {Fitzsimmons},
  {Gezari}, {Gibbs}, {Goessl}, {Goggia}, {Gourgue}, {Goldman}, {Grant},
  {Grebel}, {Hambly}, {Hasinger}, {Heavens}, {Heckman}, {Henderson}, {Henning},
  {Holman}, {Hopp}, {Ip}, {Isani}, {Jackson}, {Keyes}, {Koekemoer}, {Kotak},
  {Le}, {Liska}, {Long}, {Lucey}, {Liu}, {Martin}, {Masci}, {McLean}, {Mindel},
  {Misra}, {Morganson}, {Murphy}, {Obaika}, {Narayan}, {Nieto-Santisteban},
  {Norberg}, {Peacock}, {Pier}, {Postman}, {Primak}, {Rae}, {Rai}, {Riess},
  {Riffeser}, {Rix}, {R{\"o}ser}, {Russel}, {Rutz}, {Schilbach}, {Schultz},
  {Scolnic}, {Strolger}, {Szalay}, {Seitz}, {Small}, {Smith}, {Soderblom},
  {Taylor}, {Thomson}, {Taylor}, {Thakar}, {Thiel}, {Thilker}, {Unger},
  {Urata}, {Valenti}, {Wagner}, {Walder}, {Walter}, {Watters}, {Werner},
  {Wood-Vasey}, \& {Wyse}}]{2016arXiv161205560C}
{Chambers}, K.~C., {Magnier}, E.~A., {Metcalfe}, N., {et~al.} 2016, arXiv
  e-prints, arXiv:1612.05560, \dodoi{10.48550/arXiv.1612.05560}

\bibitem[{{Croom} {et~al.}(2004){Croom}, {Smith}, {Boyle}, {Shanks}, {Miller},
  {Outram}, \& {Loaring}}]{2004MNRAS.349.1397C}
{Croom}, S.~M., {Smith}, R.~J., {Boyle}, B.~J., {et~al.} 2004, \mnras, 349,
  1397, \dodoi{10.1111/j.1365-2966.2004.07619.x}

\bibitem[{{Croom} {et~al.}(2005){Croom}, {Boyle}, {Shanks}, {Smith}, {Miller},
  {Outram}, {Loaring}, {Hoyle}, \& {da {\^A}ngela}}]{2005MNRAS.356..415C}
{Croom}, S.~M., {Boyle}, B.~J., {Shanks}, T., {et~al.} 2005, \mnras, 356, 415,
  \dodoi{10.1111/j.1365-2966.2004.08379.x}

\bibitem[{{Cui} {et~al.}(2012){Cui}, {Zhao}, {Chu}, {Li}, {Li}, {Zhang}, {Su},
  {Yao}, {Wang}, {Xing}, {Li}, {Zhu}, {Wang}, {Gu}, {Luo}, {Xu}, {Zhang},
  {Liu}, {Zhang}, {Yang}, {Cao}, {Chen}, {Chen}, {Chen}, {Chen}, {Chu}, {Feng},
  {Gong}, {Hou}, {Hu}, {Hu}, {Hu}, {Jia}, {Jiang}, {Jiang}, {Jiang}, {Jin},
  {Li}, {Li}, {Li}, {Liu}, {Liu}, {Lu}, {Mao}, {Men}, {Qi}, {Qi}, {Shi},
  {Tang}, {Tao}, {Wang}, {Wang}, {Wang}, {Wang}, {Wang}, {Wang}, {Wang},
  {Wang}, {Wang}, {Wang}, {Wang}, {Wang}, {Xu}, {Xu}, {Yang}, {Yu}, {Yuan},
  {Yuan}, {Zhai}, {Zhang}, {Zhang}, {Zhang}, {Zhao}, {Zhou}, {Zhou}, {Zhu}, \&
  {Zou}}]{2012RAA....12.1197C}
{Cui}, X.-Q., {Zhao}, Y.-H., {Chu}, Y.-Q., {et~al.} 2012, Research in Astronomy
  and Astrophysics, 12, 1197, \dodoi{10.1088/1674-4527/12/9/003}

\bibitem[{{Cutri} {et~al.}(2013){Cutri}, {Wright}, {Conrow}, {Fowler},
  {Eisenhardt}, {Grillmair}, {Kirkpatrick}, {Masci}, {McCallon}, {Wheelock},
  {Fajardo-Acosta}, {Yan}, {Benford}, {Harbut}, {Jarrett}, {Lake}, {Leisawitz},
  {Ressler}, {Stanford}, {Tsai}, {Liu}, {Helou}, {Mainzer}, {Gettings},
  {Gonzalez}, {Hoffman}, {Marsh}, {Padgett}, {Skrutskie}, {Beck}, {Papin}, \&
  {Wittman}}]{2013wise.rept....1C}
{Cutri}, R.~M., {Wright}, E.~L., {Conrow}, T., {et~al.} 2013, {Explanatory
  Supplement to the AllWISE Data Release Products}, Explanatory Supplement to
  the AllWISE Data Release Products, by R. M. Cutri et al.

\bibitem[{{DESI Collaboration} {et~al.}(2024){DESI Collaboration}, {Adame},
  {Aguilar}, {Ahlen}, {Alam}, {Aldering}, {Alexander}, {Alfarsy}, {Allende
  Prieto}, {Alvarez}, {Alves}, {Anand}, {Andrade-Oliveira}, {Armengaud},
  {Asorey}, {Avila}, {Aviles}, {Bailey}, {Balaguera-Antol{\'\i}nez},
  {Ballester}, {Baltay}, {Bault}, {Bautista}, {Behera}, {Beltran}, {BenZvi},
  {Beraldo e Silva}, {Bermejo-Climent}, {Berti}, {Besuner}, {Beutler},
  {Bianchi}, {Blake}, {Blum}, {Bolton}, {Brieden}, {Brodzeller}, {Brooks},
  {Brown}, {Buckley-Geer}, {Burtin}, {Cabayol-Garcia}, {Cai}, {Canning},
  {Cardiel-Sas}, {Carnero Rosell}, {Castander}, {Cervantes-Cota}, {Chabanier},
  {Chaussidon}, {Chaves-Montero}, {Chen}, {Chen}, {Chuang}, {Claybaugh},
  {Cole}, {Cooper}, {Cuceu}, {Davis}, {Dawson}, {de Belsunce}, {de la Cruz},
  {de la Macorra}, {Della Costa}, {de Mattia}, {Demina}, {Demirbozan},
  {DeRose}, {Dey}, {Dey}, {Dhungana}, {Ding}, {Ding}, {Doel}, {Doshi},
  {Douglass}, {Edge}, {Eftekharzadeh}, {Eisenstein}, {Elliott}, {Ereza},
  {Escoffier}, {Fagrelius}, {Fan}, {Fanning}, {Fawcett}, {Ferraro}, {Flaugher},
  {Font-Ribera}, {Forero-Romero}, {Forero-S{\'a}nchez}, {Frenk},
  {G{\"a}nsicke}, {Garc{\'\i}a}, {Garc{\'\i}a-Bellido}, {Garcia-Quintero},
  {Garrison}, {Gil-Mar{\'\i}n}, {Golden-Marx}, {Gontcho A Gontcho},
  {Gonzalez-Morales}, {Gonzalez-Perez}, {Gordon}, {Graur}, {Green}, {Gruen},
  {Guy}, {Hadzhiyska}, {Hahn}, {Han}, {Hanif}, {Herrera-Alcantar}, {Honscheid},
  {Hou}, {Howlett}, {Huterer}, {Ir{\v{s}}i{\v{c}}}, {Ishak}, {Jacques}, {Jana},
  {Jiang}, {Jimenez}, {Jing}, {Joudaki}, {Joyce}, {Jullo}, {Juneau},
  {Kara{\c{c}}ayl{\i}}, {Karim}, {Kehoe}, {Kent}, {Khederlarian}, {Kim},
  {Kirkby}, {Kisner}, {Kitaura}, {Kizhuprakkat}, {Kneib}, {Koposov},
  {Kov{\'a}cs}, {Kremin}, {Krolewski}, {L'Huillier}, {Lahav}, {Lambert},
  {Lamman}, {Lan}, {Landriau}, {Lang}, {Lange}, {Lasker}, {Leauthaud}, {Le
  Guillou}, {Levi}, {Li}, {Linder}, {Lyons}, {Magneville}, {Manera}, {Manser},
  {Margala}, {Martini}, {McDonald}, {Medina}, {Medina-Varela}, {Meisner},
  {Mena-Fern{\'a}ndez}, {Meneses-Rizo}, {Mezcua}, {Miquel}, {Montero-Camacho},
  {Moon}, {Moore}, {Moustakas}, {Mueller}, {Mundet}, {Mu{\~n}oz-Guti{\'e}rrez},
  {Myers}, {Nadathur}, {Napolitano}, {Neveux}, {Newman}, {Nie}, {Nikutta},
  {Niz}, {Norberg}, {Noriega}, {Paillas}, {Palanque-Delabrouille}, {Palmese},
  {Pan}, {Parkinson}, {Penmetsa}, {Percival}, {P{\'e}rez-Fern{\'a}ndez},
  {P{\'e}rez-R{\`a}fols}, {Pieri}, {Poppett}, {Porredon}, {Pothier}, {Prada},
  {Pucha}, {Raichoor}, {Ram{\'\i}rez-P{\'e}rez}, {Ramirez-Solano},
  {Rashkovetskyi}, {Ravoux}, {Rocher}, {Rockosi}, {Ross}, {Rossi}, {Ruggeri},
  {Ruhlmann-Kleider}, {Sabiu}, {Said}, {Saintonge}, {Samushia}, {Sanchez},
  {Saulder}, {Schaan}, {Schlafly}, {Schlegel}, {Scholte}, {Schubnell}, {Seo},
  {Shafieloo}, {Sharples}, {Sheu}, {Silber}, {Sinigaglia}, {Siudek}, {Slepian},
  {Smith}, {Soumagnac}, {Sprayberry}, {Stephey}, {Su{\'a}rez-P{\'e}rez}, {Sun},
  {Tan}, {Tarl{\'e}}, {Tojeiro}, {Ure{\~n}a-L{\'o}pez}, {Vaisakh}, {Valcin},
  {Valdes}, {Valluri}, {Vargas-Maga{\~n}a}, {Variu}, {Verde}, {Walther},
  {Wang}, {Wang}, {Weaver}, {Weaverdyck}, {Wechsler}, {White}, {Xie}, {Yang},
  {Y{\`e}che}, {Yu}, {Yuan}, {Zhang}, {Zhang}, {Zhao}, {Zheng}, {Zhou}, {Zhou},
  {Zou}, {Zou}, \& {Zu}}]{2024AJ....168...58D}
{DESI Collaboration}, {Adame}, A.~G., {Aguilar}, J., {et~al.} 2024, \aj, 168,
  58, \dodoi{10.3847/1538-3881/ad3217}

\bibitem[{{Ding} {et~al.}(2024){Ding}, {Liao}, {Wu}, {Qi}, \&
  {Tang}}]{2024A&A...691A..81D}
{Ding}, Y., {Liao}, S., {Wu}, Q., {Qi}, Z., \& {Tang}, Z. 2024, \aap, 691, A81,
  \dodoi{10.1051/0004-6361/202450967}

\bibitem[{{Dong} {et~al.}(2018){Dong}, {Wu}, {Ai}, {Yang}, {Yang}, {Wang},
  {Zhang}, {Luo}, {Xu}, {Yuan}, {Zhang}, {Wang}, {Wang}, {Li}, {Zuo}, {Hou},
  {Guo}, {Kong}, {Chen}, {Wu}, {Yang}, \& {Yang}}]{2018AJ....155..189D}
{Dong}, X.~Y., {Wu}, X.-B., {Ai}, Y.~L., {et~al.} 2018, \aj, 155, 189,
  \dodoi{10.3847/1538-3881/aab5ae}

\bibitem[{{Falomo} {et~al.}(2005){Falomo}, {Kotilainen}, {Scarpa}, \&
  {Treves}}]{2005A&A...434..469F}
{Falomo}, R., {Kotilainen}, J.~K., {Scarpa}, R., \& {Treves}, A. 2005, \aap,
  434, 469, \dodoi{10.1051/0004-6361:20041894}

\bibitem[{{Fan} {et~al.}(2006){Fan}, {Strauss}, {Becker}, {White}, {Gunn},
  {Knapp}, {Richards}, {Schneider}, {Brinkmann}, \&
  {Fukugita}}]{2006AJ....132..117F}
{Fan}, X., {Strauss}, M.~A., {Becker}, R.~H., {et~al.} 2006, \aj, 132, 117,
  \dodoi{10.1086/504836}

\bibitem[{{Flesch}(2023)}]{2023OJAp....6E..49F}
{Flesch}, E.~W. 2023, The Open Journal of Astrophysics, 6, 49,
  \dodoi{10.21105/astro.2308.01505}

\bibitem[{Fraunhofer(1817)}]{fraunhofer1817bestimmung}
Fraunhofer, J. 1817, Annalen der Physik, 56, 264

\bibitem[{{Fu} {et~al.}(2021){Fu}, {Wu}, {Yang}, {Brown}, {Feng}, {Ma}, \&
  {Li}}]{2021ApJS..254....6F}
{Fu}, Y., {Wu}, X.-B., {Yang}, Q., {et~al.} 2021, \apjs, 254, 6,
  \dodoi{10.3847/1538-4365/abe85e}

\bibitem[{{Fu} {et~al.}(2022){Fu}, {Wu}, {Jiang}, {Zhang}, {Huo}, {Ai}, {Yang},
  {Ma}, {Feng}, {Joshi}, {Hon}, {Wolf}, {Li}, {Jin}, {Yao}, {Pang}, {Wang},
  {Lu}, {Wang}, {Zheng}, {Xu}, {Yu}, {Lun}, \& {Zuo}}]{2022ApJS..261...32F}
{Fu}, Y., {Wu}, X.-B., {Jiang}, L., {et~al.} 2022, \apjs, 261, 32,
  \dodoi{10.3847/1538-4365/ac7f3e}

\bibitem[{{Gaia Collaboration}(2022)}]{Gaia_DR3_2022}
{Gaia Collaboration}. 2022, {VizieR Online Data Catalog: Gaia DR3 Part 1. Main
  source (Gaia Collaboration, 2022)}, VizieR On-line Data Catalog: I/355.
  Originally published in: Astron. Astrophys., in prep. (2022),
  \dodoi{10.26093/cds/vizier.1355}

\bibitem[{{Gaia Collaboration} {et~al.}(2018){Gaia Collaboration}, {Mignard},
  {Klioner}, {Lindegren}, {Hern{\'a}ndez}, {Bastian}, {Bombrun}, {Hobbs},
  {Lammers}, {Michalik}, {Ramos-Lerate}, {Biermann},
  {Fern{\'a}ndez-Hern{\'a}ndez}, {Geyer}, {Hilger}, {Siddiqui},
  {Steidelm{\"u}ller}, {Babusiaux}, {Barache}, {Lambert}, {Andrei}, {Bourda},
  {Charlot}, {Brown}, {Vallenari}, {Prusti}, {de Bruijne}, {Bailer-Jones},
  {Evans}, {Eyer}, {Jansen}, {Jordi}, {Luri}, {Panem}, {Pourbaix}, {Randich},
  {Sartoretti}, {Soubiran}, {van Leeuwen}, {Walton}, {Arenou}, {Cropper},
  {Drimmel}, {Katz}, {Lattanzi}, {Bakker}, {Cacciari}, {Casta{\~n}eda},
  {Chaoul}, {Cheek}, {De Angeli}, {Fabricius}, {Guerra}, {Holl}, {Masana},
  {Messineo}, {Mowlavi}, {Nienartowicz}, {Panuzzo}, {Portell}, {Riello},
  {Seabroke}, {Tanga}, {Th{\'e}venin}, {Gracia-Abril}, {Comoretto},
  {Garcia-Reinaldos}, {Teyssier}, {Altmann}, {Andrae}, {Audard},
  {Bellas-Velidis}, {Benson}, {Berthier}, {Blomme}, {Burgess}, {Busso},
  {Carry}, {Cellino}, {Clementini}, {Clotet}, {Creevey}, {Davidson}, {De
  Ridder}, {Delchambre}, {Dell'Oro}, {Ducourant}, {Fouesneau}, {Fr{\'e}mat},
  {Galluccio}, {Garc{\'\i}a-Torres}, {Gonz{\'a}lez-N{\'u}{\~n}ez},
  {Gonz{\'a}lez-Vidal}, {Gosset}, {Guy}, {Halbwachs}, {Hambly}, {Harrison},
  {Hestroffer}, {Hodgkin}, {Hutton}, {Jasniewicz}, {Jean-Antoine-Piccolo},
  {Jordan}, {Korn}, {Krone-Martins}, {Lanzafame}, {Lebzelter}, {L{\"o}ffler},
  {Manteiga}, {Marrese}, {Mart{\'\i}n-Fleitas}, {Moitinho}, {Mora}, {Muinonen},
  {Osinde}, {Pancino}, {Pauwels}, {Petit}, {Recio-Blanco}, {Richards},
  {Rimoldini}, {Robin}, {Sarro}, {Siopis}, {Smith}, {Sozzetti}, {S{\"u}veges},
  {Torra}, {van Reeven}, {Abbas}, {Abreu Aramburu}, {Accart}, {Aerts},
  {Altavilla}, {{\'A}lvarez}, {Alvarez}, {Alves}, {Anderson}, {Anglada Varela},
  {Antiche}, {Antoja}, {Arcay}, {Astraatmadja}, {Bach}, {Baker},
  {Balaguer-N{\'u}{\~n}ez}, {Balm}, {Barata}, {Barbato}, {Barblan}, {Barklem},
  {Barrado}, {Barros}, {Barstow}, {Bartholom{\'e} Mu{\~n}oz}, {Bassilana},
  {Becciani}, {Bellazzini}, {Berihuete}, {Bertone}, {Bianchi}, {Bienaym{\'e}},
  {Blanco-Cuaresma}, {Boch}, {Boeche}, {Borrachero}, {Bossini}, {Bouquillon},
  {Bragaglia}, {Bramante}, {Breddels}, {Bressan}, {Brouillet},
  {Br{\"u}semeister}, {Brugaletta}, {Bucciarelli}, {Burlacu}, {Busonero},
  {Butkevich}, {Buzzi}, {Caffau}, {Cancelliere}, {Cannizzaro}, {Cantat-Gaudin},
  {Carballo}, {Carlucci}, {Carrasco}, {Casamiquela}, {Castellani},
  {Castro-Ginard}, {Chemin}, {Chiavassa}, {Cocozza}, {Costigan}, {Cowell},
  {Crifo}, {Crosta}, {Crowley}, {Cuypers}, {Dafonte}, {Damerdji}, {Dapergolas},
  {David}, {David}, {de Laverny}, {De Luise}, {De March}, {de Souza}, {de
  Torres}, {Debosscher}, {del Pozo}, {Delbo}, {Delgado}, {Delgado}, {Diakite},
  {Diener}, {Distefano}, {Dolding}, {Drazinos}, {Dur{\'a}n}, {Edvardsson},
  {Enke}, {Eriksson}, {Esquej}, {Eynard Bontemps}, {Fabre}, {Fabrizio},
  {Faigler}, {Falc{\~a}o}, {Farr{\`a}s Casas}, {Federici}, {Fedorets},
  {Fernique}, {Figueras}, {Filippi}, {Findeisen}, {Fonti}, {Fraile}, {Fraser},
  {Fr{\'e}zouls}, {Gai}, {Galleti}, {Garabato}, {Garc{\'\i}a-Sedano},
  {Garofalo}, {Garralda}, {Gavel}, {Gavras}, {Gerssen}, {Giacobbe}, {Gilmore},
  {Girona}, {Giuffrida}, {Glass}, {Gomes}, {Granvik}, {Gueguen}, {Guerrier},
  {Guiraud}, {Guti{\'e}}, {Haigron}, {Hatzidimitriou}, {Hauser}, {Haywood},
  {Heiter}, {Helmi}, {Heu}, {Hofmann}, {Holland}, {Huckle}, {Hypki}, {Icardi},
  {Jan{\ss}en}, {Jevardat de Fombelle}, {Jonker}, {Juh{\'a}sz}, {Julbe},
  {Karampelas}, {Kewley}, {Klar}, {Kochoska}, {Kohley}, {Kolenberg},
  {Kontizas}, {Kontizas}, {Koposov}, {Kordopatis}, {Kostrzewa-Rutkowska},
  {Koubsky}, {Lanza}, {Lasne}, {Lavigne}, {Le Fustec}, {Le Poncin-Lafitte},
  {Lebreton}, {Leccia}, {Leclerc}, {Lecoeur-Taibi}, {Lenhardt}, {Leroux},
  {Liao}, {Licata}, {Lindstr{\o}m}, {Lister}, {Livanou}, {Lobel}, {L{\'o}pez},
  {Managau}, {Mann}, {Mantelet}, {Marchal}, {Marchant}, {Marconi}, {Marinoni},
  {Marschalk{\'o}}, {Marshall}, {Martino}, {Marton}, {Mary}, {Massari},
  {Matijevi{\v{c}}}, {Mazeh}, {McMillan}, {Messina}, {Millar}, {Molina},
  {Molinaro}, {Moln{\'a}r}, {Montegriffo}, {Mor}, {Morbidelli}, {Morel},
  {Morris}, {Mulone}, {Muraveva}, {Musella}, {Nelemans}, {Nicastro}, {Noval},
  {O'Mullane}, {Ord{\'e}novic}, {Ord{\'o}{\~n}ez-Blanco}, {Osborne}, {Pagani},
  {Pagano}, {Pailler}, {Palacin}, {Palaversa}, {Panahi}, {Pawlak},
  {Piersimoni}, {Pineau}, {Plachy}, {Plum}, {Poggio}, {Poujoulet},
  {Pr{\v{s}}a}, {Pulone}, {Racero}, {Ragaini}, {Rambaux}, {Regibo},
  {Reyl{\'e}}, {Riclet}, {Ripepi}, {Riva}, {Rivard}, {Rixon}, {Roegiers},
  {Roelens}, {Romero-G{\'o}mez}, {Rowell}, {Royer}, {Ruiz-Dern}, {Sadowski},
  {Sagrist{\`a} Sell{\'e}s}, {Sahlmann}, {Salgado}, {Salguero}, {Sanna},
  {Santana-Ros}, {Sarasso}, {Savietto}, {Schultheis}, {Sciacca}, {Segol},
  {Segovia}, {S{\'e}gransan}, {Shih}, {Siltala}, {Silva}, {Smart}, {Smith},
  {Solano}, {Solitro}, {Sordo}, {Soria Nieto}, {Souchay}, {Spagna}, {Spoto},
  {Stampa}, {Steele}, {Stephenson}, {Stoev}, {Suess}, {Surdej}, {Szabados},
  {Szegedi-Elek}, {Tapiador}, {Taris}, {Tauran}, {Taylor}, {Teixeira},
  {Terrett}, {Teyssandier}, {Thuillot}, {Titarenko}, {Torra Clotet}, {Turon},
  {Ulla}, {Utrilla}, {Uzzi}, {Vaillant}, {Valentini}, {Valette}, {van Elteren},
  {Van Hemelryck}, {van Leeuwen}, {Vaschetto}, {Vecchiato}, {Veljanoski},
  {Viala}, {Vicente}, {Vogt}, {von Essen}, {Voss}, {Votruba}, {Voutsinas},
  {Walmsley}, {Weiler}, {Wertz}, {Wevers}, {Wyrzykowski}, {Yoldas},
  {{\v{Z}}erjal}, {Ziaeepour}, {Zorec}, {Zschocke}, {Zucker}, {Zurbach}, \&
  {Zwitter}}]{2018A&A...616A..14G}
{Gaia Collaboration}, {Mignard}, F., {Klioner}, S.~A., {et~al.} 2018, \aap,
  616, A14, \dodoi{10.1051/0004-6361/201832916}

\bibitem[{{Gaia Collaboration} {et~al.}(2021){Gaia Collaboration}, {Klioner},
  {Mignard}, {Lindegren}, {Bastian}, {McMillan}, {Hern{\'a}ndez}, {Hobbs},
  {Ramos-Lerate}, {Biermann}, {Bombrun}, {de Torres}, {Gerlach}, {Geyer},
  {Hilger}, {Lammers}, {Steidelm{\"u}ller}, {Stephenson}, {Brown}, {Vallenari},
  {Prusti}, {de Bruijne}, {Babusiaux}, {Creevey}, {Evans}, {Eyer}, {Hutton},
  {Jansen}, {Jordi}, {Luri}, {Panem}, {Pourbaix}, {Randich}, {Sartoretti},
  {Soubiran}, {Walton}, {Arenou}, {Bailer-Jones}, {Cropper}, {Drimmel}, {Katz},
  {Lattanzi}, {van Leeuwen}, {Bakker}, {Casta{\~n}eda}, {De Angeli},
  {Ducourant}, {Fabricius}, {Fouesneau}, {Fr{\'e}mat}, {Guerra}, {Guerrier},
  {Guiraud}, {Jean-Antoine Piccolo}, {Masana}, {Messineo}, {Mowlavi},
  {Nicolas}, {Nienartowicz}, {Pailler}, {Panuzzo}, {Riclet}, {Roux},
  {Seabroke}, {Sordo}, {Tanga}, {Th{\'e}venin}, {Gracia-Abril}, {Portell},
  {Teyssier}, {Altmann}, {Andrae}, {Bellas-Velidis}, {Benson}, {Berthier},
  {Blomme}, {Brugaletta}, {Burgess}, {Busso}, {Carry}, {Cellino}, {Cheek},
  {Clementini}, {Damerdji}, {Davidson}, {Delchambre}, {Dell'Oro},
  {Fern{\'a}ndez-Hern{\'a}ndez}, {Galluccio}, {Garc{\'\i}a-Lario},
  {Garcia-Reinaldos}, {Gonz{\'a}lez-N{\'u}{\~n}ez}, {Gosset}, {Haigron},
  {Halbwachs}, {Hambly}, {Harrison}, {Hatzidimitriou}, {Heiter}, {Hestroffer},
  {Hodgkin}, {Holl}, {Jan{\ss}en}, {Jevardat de Fombelle}, {Jordan},
  {Krone-Martins}, {Lanzafame}, {L{\"o}ffler}, {Lorca}, {Manteiga}, {Marchal},
  {Marrese}, {Moitinho}, {Mora}, {Muinonen}, {Osborne}, {Pancino}, {Pauwels},
  {Recio-Blanco}, {Richards}, {Riello}, {Rimoldini}, {Robin}, {Roegiers},
  {Rybizki}, {Sarro}, {Siopis}, {Smith}, {Sozzetti}, {Ulla}, {Utrilla}, {van
  Leeuwen}, {van Reeven}, {Abbas}, {Abreu Aramburu}, {Accart}, {Aerts},
  {Aguado}, {Ajaj}, {Altavilla}, {{\'A}lvarez}, {{\'A}lvarez Cid-Fuentes},
  {Alves}, {Anderson}, {Anglada Varela}, {Antoja}, {Audard}, {Baines}, {Baker},
  {Balaguer-N{\'u}{\~n}ez}, {Balbinot}, {Balog}, {Barache}, {Barbato},
  {Barros}, {Barstow}, {Bartolom{\'e}}, {Bassilana}, {Bauchet},
  {Baudesson-Stella}, {Becciani}, {Bellazzini}, {Bernet}, {Bertone}, {Bianchi},
  {Blanco-Cuaresma}, {Boch}, {Bossini}, {Bouquillon}, {Bramante}, {Breedt},
  {Bressan}, {Brouillet}, {Bucciarelli}, {Burlacu}, {Busonero}, {Butkevich},
  {Buzzi}, {Caffau}, {Cancelliere}, {C{\'a}novas}, {Cantat-Gaudin}, {Carballo},
  {Carlucci}, {Carnerero}, {Carrasco}, {Casamiquela}, {Castellani},
  {Castro-Ginard}, {Castro Sampol}, {Chaoul}, {Charlot}, {Chemin}, {Chiavassa},
  {Comoretto}, {Cooper}, {Cornez}, {Cowell}, {Crifo}, {Crosta}, {Crowley},
  {Dafonte}, {Dapergolas}, {David}, {David}, {de Laverny}, {De Luise}, {De
  March}, {De Ridder}, {de Souza}, {de Teodoro}, {del Peloso}, {del Pozo},
  {Delgado}, {Delgado}, {Delisle}, {Di Matteo}, {Diakite}, {Diener},
  {Distefano}, {Dolding}, {Eappachen}, {Enke}, {Esquej}, {Fabre}, {Fabrizio},
  {Faigler}, {Fedorets}, {Fernique}, {Fienga}, {Figueras}, {Fouron},
  {Fragkoudi}, {Fraile}, {Franke}, {Gai}, {Garabato}, {Garcia-Gutierrez},
  {Garc{\'\i}a-Torres}, {Garofalo}, {Gavras}, {Giacobbe}, {Gilmore}, {Girona},
  {Giuffrida}, {Gomez}, {Gonzalez-Santamaria}, {Gonz{\'a}lez-Vidal}, {Granvik},
  {Guti{\'e}rrez-S{\'a}nchez}, {Guy}, {Hauser}, {Haywood}, {Helmi}, {Hidalgo},
  {H{\l}adczuk}, {Holland}, {Huckle}, {Jasniewicz}, {Jonker}, {Juaristi
  Campillo}, {Julbe}, {Karbevska}, {Kervella}, {Khanna}, {Kochoska},
  {Kordopatis}, {Korn}, {Kostrzewa-Rutkowska}, {Kruszy{\'n}ska}, {Lambert},
  {Lanza}, {Lasne}, {Le Campion}, {Le Fustec}, {Lebreton}, {Lebzelter},
  {Leccia}, {Leclerc}, {Lecoeur-Taibi}, {Liao}, {Licata}, {Lindstr{\o}m},
  {Lister}, {Livanou}, {Lobel}, {Madrero Pardo}, {Managau}, {Mann}, {Marchant},
  {Marconi}, {Marcos Santos}, {Marinoni}, {Marocco}, {Marshall}, {Martin Polo},
  {Mart{\'\i}n-Fleitas}, {Masip}, {Massari}, {Mastrobuono-Battisti}, {Mazeh},
  {Messina}, {Michalik}, {Millar}, {Mints}, {Molina}, {Molinaro}, {Moln{\'a}r},
  {Montegriffo}, {Mor}, {Morbidelli}, {Morel}, {Morris}, {Mulone}, {Munoz},
  {Muraveva}, {Murphy}, {Musella}, {Noval}, {Ord{\'e}novic}, {Orr{\`u}},
  {Osinde}, {Pagani}, {Pagano}, {Palaversa}, {Palicio}, {Panahi}, {Pawlak},
  {Pe{\~n}alosa Esteller}, {Penttil{\"a}}, {Piersimoni}, {Pineau}, {Plachy},
  {Plum}, {Poggio}, {Poretti}, {Poujoulet}, {Pr{\v{s}}a}, {Pulone}, {Racero},
  {Ragaini}, {Rainer}, {Raiteri}, {Rambaux}, {Ramos}, {Re Fiorentin}, {Regibo},
  {Reyl{\'e}}, {Ripepi}, {Riva}, {Rixon}, {Robichon}, {Robin}, {Roelens},
  {Rohrbasser}, {Romero-G{\'o}mez}, {Rowell}, {Royer}, {Rybicki}, {Sadowski},
  {Sagrist{\`a} Sell{\'e}s}, {Sahlmann}, {Salgado}, {Salguero}, {Samaras},
  {Sanchez Gimenez}, {Sanna}, {Santove{\~n}a}, {Sarasso}, {Schultheis},
  {Sciacca}, {Segol}, {Segovia}, {S{\'e}gransan}, {Semeux}, {Siddiqui},
  {Siebert}, {Siltala}, {Slezak}, {Smart}, {Solano}, {Solitro}, {Souami},
  {Souchay}, {Spagna}, {Spoto}, {Steele}, {S{\"u}veges}, {Szabados},
  {Szegedi-Elek}, {Taris}, {Tauran}, {Taylor}, {Teixeira}, {Thuillot},
  {Tonello}, {Torra}, {Torra}, {Turon}, {Unger}, {Vaillant}, {van Dillen},
  {Vanel}, {Vecchiato}, {Viala}, {Vicente}, {Voutsinas}, {Weiler}, {Wevers},
  {Wyrzykowski}, {Yoldas}, {Yvard}, {Zhao}, {Zorec}, {Zucker}, {Zurbach}, \&
  {Zwitter}}]{2021A&A...649A...9G}
{Gaia Collaboration}, {Klioner}, S.~A., {Mignard}, F., {et~al.} 2021, \aap,
  649, A9, \dodoi{10.1051/0004-6361/202039734}

\bibitem[{{Gaia Collaboration} {et~al.}(2022){Gaia Collaboration}, {Klioner},
  {Lindegren}, {Mignard}, {Hern{\'a}ndez}, {Ramos-Lerate}, {Bastian},
  {Biermann}, {Bombrun}, {de Torres}, {Gerlach}, {Geyer}, {Hilger}, {Hobbs},
  {Lammers}, {McMillan}, {Steidelm{\"u}ller}, {Teyssier}, {Raiteri},
  {Bartolom{\'e}}, {Bernet}, {Casta{\~n}eda}, {Clotet}, {Davidson},
  {Fabricius}, {Garralda Torres}, {Gonz{\'a}lez-Vidal}, {Portell}, {Rowell},
  {Torra}, {Torra}, {Brown}, {Vallenari}, {Prusti}, {de Bruijne}, {Arenou},
  {Babusiaux}, {Creevey}, {Ducourant}, {Evans}, {Eyer}, {Guerra}, {Hutton},
  {Jordi}, {Luri}, {Panem}, {Pourbaix}, {Randich}, {Sartoretti}, {Soubiran},
  {Tanga}, {Walton}, {Bailer-Jones}, {Drimmel}, {Jansen}, {Katz}, {Lattanzi},
  {van Leeuwen}, {Bakker}, {Cacciari}, {De Angeli}, {Fouesneau}, {Fr{\'e}mat},
  {Galluccio}, {Guerrier}, {Heiter}, {Masana}, {Messineo}, {Mowlavi},
  {Nicolas}, {Nienartowicz}, {Pailler}, {Panuzzo}, {Riclet}, {Roux},
  {Seabroke}, {Sordo}, {Th{\'e}venin}, {Gracia-Abril}, {Altmann}, {Andrae},
  {Audard}, {Bellas-Velidis}, {Benson}, {Berthier}, {Blomme}, {Burgess},
  {Busonero}, {Busso}, {C{\'a}novas}, {Carry}, {Cellino}, {Cheek},
  {Clementini}, {Damerdji}, {de Teodoro}, {Nu{\~n}ez Campos}, {Delchambre},
  {Dell'Oro}, {Esquej}, {Fern{\'a}ndez-Hern{\'a}ndez}, {Fraile}, {Garabato},
  {Garc{\'\i}a-Lario}, {Gosset}, {Haigron}, {Halbwachs}, {Hambly}, {Harrison},
  {Hestroffer}, {Hodgkin}, {Holl}, {Jan{\ss}en}, {Jevardat de Fombelle},
  {Jordan}, {Krone-Martins}, {Lanzafame}, {L{\"o}ffler}, {Marchal}, {Marrese},
  {Moitinho}, {Muinonen}, {Osborne}, {Pancino}, {Pauwels}, {Recio-Blanco},
  {Reyl{\'e}}, {Riello}, {Rimoldini}, {Roegiers}, {Rybizki}, {Sarro}, {Siopis},
  {Smith}, {Sozzetti}, {Utrilla}, {van Leeuwen}, {Abbas}, {{\'A}brah{\'a}m},
  {Abreu Aramburu}, {Aerts}, {Aguado}, {Ajaj}, {Aldea-Montero}, {Altavilla},
  {{\'A}lvarez}, {Alves}, {Anderson}, {Anglada Varela}, {Antoja}, {Baines},
  {Baker}, {Balaguer-N{\'u}{\~n}ez}, {Balbinot}, {Balog}, {Barache}, {Barbato},
  {Barros}, {Barstow}, {Bassilana}, {Bauchet}, {Becciani}, {Bellazzini},
  {Berihuete}, {Bertone}, {Bianchi}, {Binnenfeld}, {Blanco-Cuaresma}, {Boch},
  {Bossini}, {Bouquillon}, {Bragaglia}, {Bramante}, {Breedt}, {Bressan},
  {Brouillet}, {Brugaletta}, {Bucciarelli}, {Burlacu}, {Butkevich}, {Buzzi},
  {Caffau}, {Cancelliere}, {Cantat-Gaudin}, {Carballo}, {Carlucci},
  {Carnerero}, {Carrasco}, {Casamiquela}, {Castellani}, {Castro-Ginard},
  {Chaoul}, {Charlot}, {Chemin}, {Chiaramida}, {Chiavassa}, {Chornay},
  {Comoretto}, {Contursi}, {Cooper}, {Cornez}, {Cowell}, {Crifo}, {Cropper},
  {Crosta}, {Crowley}, {Dafonte}, {Dapergolas}, {David}, {de Laverny}, {De
  Luise}, {De March}, {De Ridder}, {de Souza}, {del Peloso}, {del Pozo},
  {Delbo}, {Delgado}, {Delisle}, {Demouchy}, {Dharmawardena}, {Diakite},
  {Diener}, {Distefano}, {Dolding}, {Enke}, {Fabre}, {Fabrizio}, {Faigler},
  {Fedorets}, {Fernique}, {Fienga}, {Figueras}, {Fournier}, {Fouron},
  {Fragkoudi}, {Gai}, {Garcia-Gutierrez}, {Garcia-Reinaldos},
  {Garc{\'\i}a-Torres}, {Garofalo}, {Gavel}, {Gavras}, {Giacobbe}, {Gilmore},
  {Girona}, {Giuffrida}, {Gomel}, {Gomez}, {Gonz{\'a}lez-N{\'u}{\~n}ez},
  {Gonz{\'a}lez-Santamar{\'\i}a}, {Granvik}, {Guillout}, {Guiraud},
  {Guti{\'e}rrez-S{\'a}nchez}, {Guy}, {Hatzidimitriou}, {Hauser}, {Haywood},
  {Helmer}, {Helmi}, {Sarmiento}, {Hidalgo}, {H{\l}adczuk}, {Holland},
  {Huckle}, {Jardine}, {Jasniewicz}, {Jean-Antoine Piccolo},
  {Jim{\'e}nez-Arranz}, {Juaristi Campillo}, {Julbe}, {Karbevska}, {Kervella},
  {Khanna}, {Kordopatis}, {Korn}, {K{\'o}sp{\'a}l}, {Kostrzewa-Rutkowska},
  {Kruszy{\'n}ska}, {Kun}, {Laizeau}, {Lambert}, {Lanza}, {Lasne}, {Le
  Campion}, {Lebreton}, {Lebzelter}, {Leccia}, {Leclerc}, {Lecoeur-Taibi},
  {Liao}, {Licata}, {Lindstr{\o}m}, {Lister}, {Livanou}, {Lobel}, {Lorca},
  {Loup}, {Madrero Pardo}, {Magdaleno Romeo}, {Managau}, {Mann}, {Manteiga},
  {Marchant}, {Marconi}, {Marcos}, {Santos}, {Mar{\'\i}n Pina}, {Marinoni},
  {Marocco}, {Marshall}, {Polo}, {Mart{\'\i}n-Fleitas}, {Marton}, {Mary},
  {Masip}, {Massari}, {Mastrobuono-Battisti}, {Mazeh}, {Messina}, {Michalik},
  {Millar}, {Mints}, {Molina}, {Molinaro}, {Moln{\'a}r}, {Monari},
  {Mongui{\'o}}, {Montegriffo}, {Montero}, {Mor}, {Mora}, {Morbidelli},
  {Morel}, {Morris}, {Muraveva}, {Murphy}, {Musella}, {Nagy}, {Noval},
  {Oca{\~n}a}, {Ogden}, {Ordenovic}, {Osinde}, {Pagani}, {Pagano}, {Palaversa},
  {Palicio}, {Pallas-Quintela}, {Panahi}, {Payne-Wardenaar}, {Pe{\~n}alosa
  Esteller}, {Penttil{\"a}}, {Pichon}, {Piersimoni}, {Pineau}, {Plachy},
  {Plum}, {Poggio}, {Pr{\v{s}}a}, {Pulone}, {Racero}, {Ragaini}, {Rainer},
  {Rambaux}, {Ramos}, {Re Fiorentin}, {Regibo}, {Richards}, {Diaz}, {Ripepi},
  {Riva}, {Rix}, {Rixon}, {Robichon}, {Robin}, {Robin}, {Roelens}, {Rogues},
  {Rohrbasser}, {Romero-G{\'o}mez}, {Royer}, {Ruz Mieres}, {Rybicki},
  {Sadowski}, {S{\'a}ez N{\'u}{\~n}ez}, {Sagrist{\`a} Sell{\'e}s}, {Sahlmann},
  {Salguero}, {Samaras}, {Sanchez Gimenez}, {Sanna}, {Santove{\~n}a},
  {Sarasso}, {Schultheis}, {Sciacca}, {Segol}, {Segovia}, {S{\'e}gransan},
  {Semeux}, {Shahaf}, {Siddiqui}, {Siebert}, {Siltala}, {Silvelo}, {Slezak},
  {Slezak}, {Smart}, {Snaith}, {Solano}, {Solitro}, {Souami}, {Souchay},
  {Spagna}, {Spina}, {Spoto}, {Steele}, {Stephenson}, {S{\"u}veges}, {Surdej},
  {Szabados}, {Szegedi-Elek}, {Taris}, {Taylor}, {Teixeira}, {Tolomei},
  {Tonello}, {Torralba Elipe}, {Trabucchi}, {Tsounis}, {Turon}, {Ulla},
  {Unger}, {Vaillant}, {van Dillen}, {van Reeven}, {Vanel}, {Vecchiato},
  {Viala}, {Vicente}, {Voutsinas}, {Weiler}, {Wevers}, {Wyrzykowski}, {Yoldas},
  {Yvard}, {Zhao}, {Zorec}, {Zucker}, \& {Zwitter}}]{2022A&A...667A.148G}
{Gaia Collaboration}, {Klioner}, S.~A., {Lindegren}, L., {et~al.} 2022, \aap,
  667, A148, \dodoi{10.1051/0004-6361/202243483}

\bibitem[{{Greig} {et~al.}(2019){Greig}, {Mesinger}, \&
  {Ba{\~n}ados}}]{2019MNRAS.484.5094G}
{Greig}, B., {Mesinger}, A., \& {Ba{\~n}ados}, E. 2019, \mnras, 484, 5094,
  \dodoi{10.1093/mnras/stz230}

\bibitem[{{Hernitschek} {et~al.}(2016){Hernitschek}, {Schlafly}, {Sesar},
  {Rix}, {Hogg}, {Ivezi{\'c}}, {Grebel}, {Bell}, {Martin}, {Burgett},
  {Flewelling}, {Hodapp}, {Kaiser}, {Magnier}, {Metcalfe}, {Wainscoat}, \&
  {Waters}}]{2016ApJ...817...73H}
{Hernitschek}, N., {Schlafly}, E.~F., {Sesar}, B., {et~al.} 2016, \apj, 817,
  73, \dodoi{10.3847/0004-637X/817/1/73}

\bibitem[{{Huo} {et~al.}(2010){Huo}, {Liu}, {Yuan}, {Zhang}, {Zhao}, {Chen},
  {Bai}, {Zhang}, {Zhang}, {Garc{\'\i}a-Benito}, {Xiang}, {Yan}, {Ren}, {Sun},
  {Zhang}, {Li}, {Lu}, {Wang}, {Ni}, \& {Wang}}]{2010RAA....10..612H}
{Huo}, Z.-Y., {Liu}, X.-W., {Yuan}, H.-B., {et~al.} 2010, Research in Astronomy
  and Astrophysics, 10, 612, \dodoi{10.1088/1674-4527/10/7/002}

\bibitem[{{Huo} {et~al.}(2013){Huo}, {Liu}, {Xiang}, {Yuan}, {Huang}, {Zhang},
  {Yan}, {Bai}, {Chen}, {Chen}, {Chu}, {Chu}, {Cui}, {Du}, {Hou}, {Hu}, {Hu},
  {Jia}, {Jiang}, {Lei}, {Li}, {Li}, {Li}, {Li}, {Li}, {Li}, {Li}, {Liu},
  {Liu}, {Lu}, {Luo}, {Luo}, {Men}, {Ni}, {Qi}, {Qi}, {Shi}, {Shi}, {Sun},
  {Tang}, {Tian}, {Tu}, {Wang}, {Wang}, {Wang}, {Wang}, {Wang}, {Wang}, {Wang},
  {Wang}, {Wei}, {Wu}, {Xue}, {Yao}, {Yu}, {Yuan}, {Zhai}, {Zhang}, {Zhang},
  {Zhang}, {Zhang}, {Zhang}, {Zhang}, {Zhang}, {Zhao}, {Zhao}, {Zhao}, {Zhou},
  {Zhou}, {Zhu}, \& {Zou}}]{2013AJ....145..159H}
{Huo}, Z.-Y., {Liu}, X.-W., {Xiang}, M.-S., {et~al.} 2013, \aj, 145, 159,
  \dodoi{10.1088/0004-6256/145/6/159}

\bibitem[{{Huo} {et~al.}(2015){Huo}, {Liu}, {Xiang}, {Shi}, {Yuan}, {Huang},
  {Zhang}, {Hou}, {Wang}, \& {Yang}}]{2015RAA....15.1438H}
---. 2015, Research in Astronomy and Astrophysics, 15, 1438,
  \dodoi{10.1088/1674-4527/15/8/023}

\bibitem[{{Huo} {et~al.}(2017){Huo}, {Liu}, {Shi}, {Xiang}, {Huang}, {Yuan},
  {Zhang}, {Zhang}, {Wang}, {Wu}, {Cao}, {Zhang}, {Hou}, \&
  {Wang}}]{2017RAA....17...32H}
{Huo}, Z.-Y., {Liu}, X.-W., {Shi}, J.-R., {et~al.} 2017, Research in Astronomy
  and Astrophysics, 17, 032, \dodoi{10.1088/1674-4527/17/4/32}

\bibitem[{{Im} {et~al.}(2007){Im}, {Lee}, {Cho}, {Choi}, {Ko}, \&
  {Song}}]{2007ApJ...664...64I}
{Im}, M., {Lee}, I., {Cho}, Y., {et~al.} 2007, \apj, 664, 64,
  \dodoi{10.1086/518734}

\bibitem[{{Jin} {et~al.}(2023){Jin}, {Wu}, {Fu}, {Yao}, {Ai}, {Feng}, {He},
  {Ma}, {Pang}, {Zhu}, {Zhang}, {Yuan}, \& {Huo}}]{2023ApJS..265...25J}
{Jin}, J.-J., {Wu}, X.-B., {Fu}, Y., {et~al.} 2023, \apjs, 265, 25,
  \dodoi{10.3847/1538-4365/acaf89}

\bibitem[{{Kim} {et~al.}(2012){Kim}, {Protopapas}, {Trichas}, {Rowan-Robinson},
  {Khardon}, {Alcock}, \& {Byun}}]{2012ApJ...747..107K}
{Kim}, D.-W., {Protopapas}, P., {Trichas}, M., {et~al.} 2012, \apj, 747, 107,
  \dodoi{10.1088/0004-637X/747/2/107}

\bibitem[{{Koz{\l}owski} \& {Kochanek}(2009)}]{2009ApJ...701..508K}
{Koz{\l}owski}, S., \& {Kochanek}, C.~S. 2009, \apj, 701, 508,
  \dodoi{10.1088/0004-637X/701/1/508}

\bibitem[{{Koz{\l}owski} {et~al.}(2011){Koz{\l}owski}, {Kochanek}, \&
  {Udalski}}]{2011ApJS..194...22K}
{Koz{\l}owski}, S., {Kochanek}, C.~S., \& {Udalski}, A. 2011, \apjs, 194, 22,
  \dodoi{10.1088/0067-0049/194/2/22}

\bibitem[{{Koz{\l}owski} {et~al.}(2012){Koz{\l}owski}, {Kochanek}, {Jacyszyn},
  {Udalski}, {Szyma{\'n}ski}, {Poleski}, {Kubiak}, {Soszy{\'n}ski},
  {Pietrzy{\'n}ski}, {Wyrzykowski}, {Ulaczyk}, \&
  {Pietrukowicz}}]{2012ApJ...746...27K}
{Koz{\l}owski}, S., {Kochanek}, C.~S., {Jacyszyn}, A.~M., {et~al.} 2012, \apj,
  746, 27, \dodoi{10.1088/0004-637X/746/1/27}

\bibitem[{{Koz{\l}owski} {et~al.}(2013){Koz{\l}owski}, {Onken}, {Kochanek},
  {Udalski}, {Szyma{\'n}ski}, {Kubiak}, {Pietrzy{\'n}ski}, {Soszy{\'n}ski},
  {Wyrzykowski}, {Ulaczyk}, {Poleski}, {Pietrukowicz}, {Skowron}, {OGLE
  Collaboration}, {Meixner}, \& {Bonanos}}]{2013ApJ...775...92K}
{Koz{\l}owski}, S., {Onken}, C.~A., {Kochanek}, C.~S., {et~al.} 2013, \apj,
  775, 92, \dodoi{10.1088/0004-637X/775/2/92}

\bibitem[{{Lindegren} {et~al.}(2021){Lindegren}, {Bastian}, {Biermann},
  {Bombrun}, {de Torres}, {Gerlach}, {Geyer}, {Hern{\'a}ndez}, {Hilger},
  {Hobbs}, {Klioner}, {Lammers}, {McMillan}, {Ramos-Lerate},
  {Steidelm{\"u}ller}, {Stephenson}, \& {van Leeuwen}}]{2021A&A...649A...4L}
{Lindegren}, L., {Bastian}, U., {Biermann}, M., {et~al.} 2021, \aap, 649, A4,
  \dodoi{10.1051/0004-6361/202039653}

\bibitem[{{Luo} {et~al.}(2015){Luo}, {Zhao}, {Zhao}, {Deng}, {Liu}, {Jing},
  {Wang}, {Zhang}, {Shi}, {Cui}, {Chu}, {Li}, {Bai}, {Wu}, {Cai}, {Cao}, {Cao},
  {Carlin}, {Chen}, {Chen}, {Chen}, {Chen}, {Chen}, {Chen}, {Chen},
  {Christlieb}, {Chu}, {Cui}, {Dong}, {Du}, {Fan}, {Feng}, {Fu}, {Gao}, {Gong},
  {Gu}, {Guo}, {Han}, {He}, {Hou}, {Hou}, {Hou}, {Hu}, {Hu}, {Hu}, {Huo},
  {Jia}, {Jiang}, {Jiang}, {Jiang}, {Jin}, {Kong}, {Kong}, {Lei}, {Li}, {Li},
  {Li}, {Li}, {Li}, {Li}, {Li}, {Li}, {Li}, {Li}, {Li}, {Li}, {Liang}, {Lin},
  {Liu}, {Liu}, {Liu}, {Liu}, {Lu}, {Luo}, {Mao}, {Newberg}, {Ni}, {Qi}, {Qi},
  {Shen}, {Shi}, {Song}, {Song}, {Su}, {Su}, {Tang}, {Tao}, {Tian}, {Wang},
  {Wang}, {Wang}, {Wang}, {Wang}, {Wang}, {Wang}, {Wang}, {Wang}, {Wang},
  {Wang}, {Wang}, {Wang}, {Wang}, {Wang}, {Wang}, {Wang}, {Wang}, {Wang},
  {Wang}, {Wei}, {Wei}, {Wu}, {Wu}, {Wu}, {Wu}, {Xing}, {Xu}, {Xu}, {Xu},
  {Yan}, {Yang}, {Yang}, {Yang}, {Yang}, {Yao}, {Yu}, {Yuan}, {Yuan}, {Yuan},
  {Yuan}, {Zhai}, {Zhang}, {Zhang}, {Zhang}, {Zhang}, {Zhang}, {Zhang},
  {Zhang}, {Zhang}, {Zhao}, {Zhou}, {Zhou}, {Zhu}, {Zhu}, {Zou}, \&
  {Zuo}}]{2015RAA....15.1095L}
{Luo}, A.~L., {Zhao}, Y.-H., {Zhao}, G., {et~al.} 2015, Research in Astronomy
  and Astrophysics, 15, 1095, \dodoi{10.1088/1674-4527/15/8/002}

\bibitem[{{Lyke} {et~al.}(2020){Lyke}, {Higley}, {McLane}, {Schurhammer},
  {Myers}, {Ross}, {Dawson}, {Chabanier}, {Martini}, {Busca}, {Mas des
  Bourboux}, {Salvato}, {Streblyanska}, {Zarrouk}, {Burtin}, {Anderson},
  {Bautista}, {Bizyaev}, {Brandt}, {Brinkmann}, {Brownstein}, {Comparat},
  {Green}, {de la Macorra}, {Mu{\~n}oz Guti{\'e}rrez}, {Hou}, {Newman},
  {Palanque-Delabrouille}, {P{\^a}ris}, {Percival}, {Petitjean}, {Rich},
  {Rossi}, {Schneider}, {Smith}, {Vivek}, \& {Weaver}}]{2020ApJS..250....8L}
{Lyke}, B.~W., {Higley}, A.~N., {McLane}, J.~N., {et~al.} 2020, \apjs, 250, 8,
  \dodoi{10.3847/1538-4365/aba623}

\bibitem[{{Murga} {et~al.}(2015){Murga}, {Zhu}, {M{\'e}nard}, \&
  {Lan}}]{2015MNRAS.452..511M}
{Murga}, M., {Zhu}, G., {M{\'e}nard}, B., \& {Lan}, T.-W. 2015, \mnras, 452,
  511, \dodoi{10.1093/mnras/stv1277}

\bibitem[{{P{\^a}ris} {et~al.}(2017){P{\^a}ris}, {Petitjean}, {Ross}, {Myers},
  {Aubourg}, {Streblyanska}, {Bailey}, {Armengaud}, {Palanque-Delabrouille},
  {Y{\`e}che}, {Hamann}, {Strauss}, {Albareti}, {Bovy}, {Bizyaev}, {Niel
  Brandt}, {Brusa}, {Buchner}, {Comparat}, {Croft}, {Dwelly}, {Fan},
  {Font-Ribera}, {Ge}, {Georgakakis}, {Hall}, {Jiang}, {Kinemuchi},
  {Malanushenko}, {Malanushenko}, {McMahon}, {Menzel}, {Merloni}, {Nandra},
  {Noterdaeme}, {Oravetz}, {Pan}, {Pieri}, {Prada}, {Salvato}, {Schlegel},
  {Schneider}, {Simmons}, {Viel}, {Weinberg}, \& {Zhu}}]{2017A&A...597A..79P}
{P{\^a}ris}, I., {Petitjean}, P., {Ross}, N.~P., {et~al.} 2017, \aap, 597, A79,
  \dodoi{10.1051/0004-6361/201527999}

\bibitem[{{P{\^a}ris} {et~al.}(2018){P{\^a}ris}, {Petitjean}, {Aubourg},
  {Myers}, {Streblyanska}, {Lyke}, {Anderson}, {Armengaud}, {Bautista},
  {Blanton}, {Blomqvist}, {Brinkmann}, {Brownstein}, {Brandt}, {Burtin},
  {Dawson}, {de la Torre}, {Georgakakis}, {Gil-Mar{\'\i}n}, {Green}, {Hall},
  {Kneib}, {LaMassa}, {Le Goff}, {MacLeod}, {Mariappan}, {McGreer}, {Merloni},
  {Noterdaeme}, {Palanque-Delabrouille}, {Percival}, {Ross}, {Rossi},
  {Schneider}, {Seo}, {Tojeiro}, {Weaver}, {Weijmans}, {Y{\`e}che}, {Zarrouk},
  \& {Zhao}}]{2018A&A...613A..51P}
{P{\^a}ris}, I., {Petitjean}, P., {Aubourg}, {\'E}., {et~al.} 2018, \aap, 613,
  A51, \dodoi{10.1051/0004-6361/201732445}

\bibitem[{{Planck Collaboration} {et~al.}(2016){Planck Collaboration},
  {Aghanim}, {Ashdown}, {Aumont}, {Baccigalupi}, {Ballardini}, {Banday},
  {Barreiro}, {Bartolo}, {Basak}, {Benabed}, {Bernard}, {Bersanelli},
  {Bielewicz}, \& et~al.}]{Planck_2016}
{Planck Collaboration}, {Aghanim}, N., {Ashdown}, M., {et~al.} 2016, \aap, 596,
  A109, \dodoi{10.1051/0004-6361/201629022}

\bibitem[{{Plane} {et~al.}(2012){Plane}, {Oetjen}, {de Miranda}, {Saiz-Lopez},
  {Gausa}, \& {Williams}}]{2012JASTP..74..181P}
{Plane}, J., {Oetjen}, H., {de Miranda}, M., {et~al.} 2012, Journal of
  Atmospheric and Solar-Terrestrial Physics, 74, 181,
  \dodoi{10.1016/j.jastp.2011.10.019}

\bibitem[{{Poznanski} {et~al.}(2012){Poznanski}, {Prochaska}, \&
  {Bloom}}]{2012MNRAS.426.1465P}
{Poznanski}, D., {Prochaska}, J.~X., \& {Bloom}, J.~S. 2012, \mnras, 426, 1465,
  \dodoi{10.1111/j.1365-2966.2012.21796.x}

\bibitem[{{Reyl{\'e}} {et~al.}(2009){Reyl{\'e}}, {Marshall}, {Robin}, \&
  {Schultheis}}]{2009A&A...495..819R}
{Reyl{\'e}}, C., {Marshall}, D.~J., {Robin}, A.~C., \& {Schultheis}, M. 2009,
  \aap, 495, 819, \dodoi{10.1051/0004-6361/200811341}

\bibitem[{{Richards} {et~al.}(2006){Richards}, {Strauss}, {Fan}, {Hall},
  {Jester}, {Schneider}, {Vanden Berk}, {Stoughton}, {Anderson}, {Brunner},
  {Gray}, {Gunn}, {Ivezi{\'c}}, {Kirkland}, {Knapp}, {Loveday}, {Meiksin},
  {Pope}, {Szalay}, {Thakar}, {Yanny}, {York}, {Barentine}, {Brewington},
  {Brinkmann}, {Fukugita}, {Harvanek}, {Kent}, {Kleinman}, {Krzesi{\'n}ski},
  {Long}, {Lupton}, {Nash}, {Neilsen}, {Nitta}, {Schlegel}, \&
  {Snedden}}]{2006AJ....131.2766R}
{Richards}, G.~T., {Strauss}, M.~A., {Fan}, X., {et~al.} 2006, \aj, 131, 2766,
  \dodoi{10.1086/503559}

\bibitem[{{Richter}(2006)}]{2006RvMA...19...31R}
{Richter}, P. 2006, Reviews in Modern Astronomy, 19, 31,
  \dodoi{10.1002/9783527619030.ch2}

\bibitem[{{Rusu} {et~al.}(2016){Rusu}, {Oguri}, {Minowa}, {Iye}, {Inada},
  {Oya}, {Kayo}, {Hayano}, {Hattori}, {Saito}, {Ito}, {Pyo}, {Terada},
  {Takami}, \& {Watanabe}}]{2016MNRAS.458....2R}
{Rusu}, C.~E., {Oguri}, M., {Minowa}, Y., {et~al.} 2016, \mnras, 458, 2,
  \dodoi{10.1093/mnras/stw092}

\bibitem[{{Savage} {et~al.}(1993){Savage}, {Lu}, {Bahcall}, {Bergeron},
  {Boksenberg}, {Hartig}, {Jannuzi}, {Kirhakos}, {Lockman}, {Sargent},
  {Schneider}, {Turnshek}, {Weymann}, \& {Wolfe}}]{1993ApJ...413..116S}
{Savage}, B.~D., {Lu}, L., {Bahcall}, J.~N., {et~al.} 1993, \apj, 413, 116,
  \dodoi{10.1086/172982}

\bibitem[{{Savage} {et~al.}(2000){Savage}, {Wakker}, {Jannuzi}, {Bahcall},
  {Bergeron}, {Boksenberg}, {Hartig}, {Kirhakos}, {Murphy}, {Sargent},
  {Schneider}, {Turnshek}, \& {Wolfe}}]{2000ApJS..129..563S}
{Savage}, B.~D., {Wakker}, B., {Jannuzi}, B.~T., {et~al.} 2000, \apjs, 129,
  563, \dodoi{10.1086/313420}

\bibitem[{{Schlafly} {et~al.}(2016){Schlafly}, {Meisner}, {Stutz},
  {Kainulainen}, {Peek}, {Tchernyshyov}, {Rix}, {Finkbeiner}, {Covey}, {Green},
  {Bell}, {Burgett}, {Chambers}, {Draper}, {Flewelling}, {Hodapp}, {Kaiser},
  {Magnier}, {Martin}, {Metcalfe}, {Wainscoat}, \&
  {Waters}}]{2016ApJ...821...78S}
{Schlafly}, E.~F., {Meisner}, A.~M., {Stutz}, A.~M., {et~al.} 2016, \apj, 821,
  78, \dodoi{10.3847/0004-637X/821/2/78}

\bibitem[{{Schlegel} {et~al.}(1998){Schlegel}, {Finkbeiner}, \&
  {Davis}}]{1998ApJ...500..525S}
{Schlegel}, D.~J., {Finkbeiner}, D.~P., \& {Davis}, M. 1998, \apj, 500, 525,
  \dodoi{10.1086/305772}

\bibitem[{{Schmidt}(1963)}]{1963Natur.197.1040S}
{Schmidt}, M. 1963, \nat, 197, 1040, \dodoi{10.1038/1971040a0}

\bibitem[{{Schneider} {et~al.}(2010){Schneider}, {Richards}, {Hall}, {Strauss},
  {Anderson}, {Boroson}, {Ross}, {Shen}, {Brandt}, {Fan}, {Inada}, {Jester},
  {Knapp}, {Krawczyk}, {Thakar}, {Vanden Berk}, {Voges}, {Yanny}, {York},
  {Bahcall}, {Bizyaev}, {Blanton}, {Brewington}, {Brinkmann}, {Eisenstein},
  {Frieman}, {Fukugita}, {Gray}, {Gunn}, {Hibon}, {Ivezi{\'c}}, {Kent}, {Kron},
  {Lee}, {Lupton}, {Malanushenko}, {Malanushenko}, {Oravetz}, {Pan}, {Pier},
  {Price}, {Saxe}, {Schlegel}, {Simmons}, {Snedden}, {SubbaRao}, {Szalay}, \&
  {Weinberg}}]{2010AJ....139.2360S}
{Schneider}, D.~P., {Richards}, G.~T., {Hall}, P.~B., {et~al.} 2010, \aj, 139,
  2360, \dodoi{10.1088/0004-6256/139/6/2360}

\bibitem[{{Shen} {et~al.}(2008){Shen}, {Greene}, {Strauss}, {Richards}, \&
  {Schneider}}]{2008ApJ...680..169S}
{Shen}, Y., {Greene}, J.~E., {Strauss}, M.~A., {Richards}, G.~T., \&
  {Schneider}, D.~P. 2008, \apj, 680, 169, \dodoi{10.1086/587475}

\bibitem[{{Silk} \& {Rees}(1998)}]{1998A&A...331L...1S}
{Silk}, J., \& {Rees}, M.~J. 1998, \aap, 331, L1,
  \dodoi{10.48550/arXiv.astro-ph/9801013}

\bibitem[{{Skrutskie} {et~al.}(2006){Skrutskie}, {Cutri}, {Stiening},
  {Weinberg}, {Schneider}, {Carpenter}, {Beichman}, {Capps}, {Chester},
  {Elias}, {Huchra}, {Liebert}, {Lonsdale}, {Monet}, {Price}, {Seitzer},
  {Jarrett}, {Kirkpatrick}, {Gizis}, {Howard}, {Evans}, {Fowler}, {Fullmer},
  {Hurt}, {Light}, {Kopan}, {Marsh}, {McCallon}, {Tam}, {Van Dyk}, \&
  {Wheelock}}]{2006AJ....131.1163S}
{Skrutskie}, M.~F., {Cutri}, R.~M., {Stiening}, R., {et~al.} 2006, \aj, 131,
  1163, \dodoi{10.1086/498708}

\bibitem[{{Su} \& {Cui}(2004)}]{2004ChJAA...4....1S}
{Su}, D.-Q., \& {Cui}, X.-Q. 2004, \cjaa, 4, 1, \dodoi{10.1088/1009-9271/4/1/1}

\bibitem[{{Vanden Berk} {et~al.}(2001){Vanden Berk}, {Richards}, {Bauer},
  {Strauss}, {Schneider}, {Heckman}, {York}, {Hall}, {Fan}, {Knapp},
  {Anderson}, {Annis}, {Bahcall}, {Bernardi}, {Briggs}, {Brinkmann}, {Brunner},
  {Burles}, {Carey}, {Castander}, {Connolly}, {Crocker}, {Csabai}, {Doi},
  {Finkbeiner}, {Friedman}, {Frieman}, {Fukugita}, {Gunn}, {Hennessy},
  {Ivezi{\'c}}, {Kent}, {Kunszt}, {Lamb}, {Leger}, {Long}, {Loveday}, {Lupton},
  {Meiksin}, {Merelli}, {Munn}, {Newberg}, {Newcomb}, {Nichol}, {Owen}, {Pier},
  {Pope}, {Rockosi}, {Schlegel}, {Siegmund}, {Smee}, {Snir}, {Stoughton},
  {Stubbs}, {SubbaRao}, {Szalay}, {Szokoly}, {Tremonti}, {Uomoto}, {Waddell},
  {Yanny}, \& {Zheng}}]{2001AJ....122..549V}
{Vanden Berk}, D.~E., {Richards}, G.~T., {Bauer}, A., {et~al.} 2001, \aj, 122,
  549, \dodoi{10.1086/321167}

\bibitem[{{Wang} {et~al.}(1996){Wang}, {Su}, {Chu}, {Cui}, \&
  {Wang}}]{1996ApOpt..35.5155W}
{Wang}, S.-G., {Su}, D.-Q., {Chu}, Y.-Q., {Cui}, X., \& {Wang}, Y.-N. 1996,
  \ao, 35, 5155, \dodoi{10.1364/AO.35.005155}

\bibitem[{{Wenger} {et~al.}(2000){Wenger}, {Ochsenbein}, {Egret}, {Dubois},
  {Bonnarel}, {Borde}, {Genova}, {Jasniewicz}, {Lalo{\"e}}, {Lesteven}, \&
  {Monier}}]{2000A&AS..143....9W}
{Wenger}, M., {Ochsenbein}, F., {Egret}, D., {et~al.} 2000, \aaps, 143, 9,
  \dodoi{10.1051/aas:2000332}

\bibitem[{{Werk} {et~al.}(2024){Werk}, {Tchernyshyov}, {Bish}, {Zheng},
  {Putman}, {Peek}, \& {Schiminovich}}]{2024ApJS..273...21W}
{Werk}, J., {Tchernyshyov}, K., {Bish}, H., {et~al.} 2024, \apjs, 273, 21,
  \dodoi{10.3847/1538-4365/ad58df}

\bibitem[{{Wright} {et~al.}(2010){Wright}, {Eisenhardt}, {Mainzer}, {Ressler},
  {Cutri}, {Jarrett}, {Kirkpatrick}, {Padgett}, {McMillan}, {Skrutskie},
  {Stanford}, {Cohen}, {Walker}, {Mather}, {Leisawitz}, {Gautier}, {McLean},
  {Benford}, {Lonsdale}, {Blain}, {Mendez}, {Irace}, {Duval}, {Liu}, {Royer},
  {Heinrichsen}, {Howard}, {Shannon}, {Kendall}, {Walsh}, {Larsen}, {Cardon},
  {Schick}, {Schwalm}, {Abid}, {Fabinsky}, {Naes}, \&
  {Tsai}}]{2010AJ....140.1868W}
{Wright}, E.~L., {Eisenhardt}, P. R.~M., {Mainzer}, A.~K., {et~al.} 2010, \aj,
  140, 1868, \dodoi{10.1088/0004-6256/140/6/1868}

\bibitem[{{Yao} {et~al.}(2019){Yao}, {Wu}, {Ai}, {Yang}, {Yang}, {Dong},
  {Joshi}, {Wang}, {Feng}, {Fu}, {Hou}, {Luo}, {Kong}, {Liu}, {Zhao}, {Zhang},
  {Yuan}, \& {Shen}}]{2019ApJS..240....6Y}
{Yao}, S., {Wu}, X.-B., {Ai}, Y.~L., {et~al.} 2019, \apjs, 240, 6,
  \dodoi{10.3847/1538-4365/aaef88}

\bibitem[{{York} {et~al.}(2000){York}, {Adelman}, {Anderson}, {Anderson},
  {Annis}, {Bahcall}, {Bakken}, {Barkhouser}, {Bastian}, {Berman}, {Boroski},
  {Bracker}, {Briegel}, {Briggs}, {Brinkmann}, {Brunner}, {Burles}, {Carey},
  {Carr}, {Castander}, {Chen}, {Colestock}, {Connolly}, {Crocker}, {Csabai},
  {Czarapata}, {Davis}, {Doi}, {Dombeck}, {Eisenstein}, {Ellman}, {Elms},
  {Evans}, {Fan}, {Federwitz}, {Fiscelli}, {Friedman}, {Frieman}, {Fukugita},
  {Gillespie}, {Gunn}, {Gurbani}, {de Haas}, {Haldeman}, {Harris}, {Hayes},
  {Heckman}, {Hennessy}, {Hindsley}, {Holm}, {Holmgren}, {Huang}, {Hull},
  {Husby}, {Ichikawa}, {Ichikawa}, {Ivezi{\'c}}, {Kent}, {Kim}, {Kinney},
  {Klaene}, {Kleinman}, {Kleinman}, {Knapp}, {Korienek}, {Kron}, {Kunszt},
  {Lamb}, {Lee}, {Leger}, {Limmongkol}, {Lindenmeyer}, {Long}, {Loomis},
  {Loveday}, {Lucinio}, {Lupton}, {MacKinnon}, {Mannery}, {Mantsch}, {Margon},
  {McGehee}, {McKay}, {Meiksin}, {Merelli}, {Monet}, {Munn}, {Narayanan},
  {Nash}, {Neilsen}, {Neswold}, {Newberg}, {Nichol}, {Nicinski}, {Nonino},
  {Okada}, {Okamura}, {Ostriker}, {Owen}, {Pauls}, {Peoples}, {Peterson},
  {Petravick}, {Pier}, {Pope}, {Pordes}, {Prosapio}, {Rechenmacher}, {Quinn},
  {Richards}, {Richmond}, {Rivetta}, {Rockosi}, {Ruthmansdorfer}, {Sandford},
  {Schlegel}, {Schneider}, {Sekiguchi}, {Sergey}, {Shimasaku}, {Siegmund},
  {Smee}, {Smith}, {Snedden}, {Stone}, {Stoughton}, {Strauss}, {Stubbs},
  {SubbaRao}, {Szalay}, {Szapudi}, {Szokoly}, {Thakar}, {Tremonti}, {Tucker},
  {Uomoto}, {Vanden Berk}, {Vogeley}, {Waddell}, {Wang}, {Watanabe},
  {Weinberg}, {Yanny}, {Yasuda}, \& {SDSS Collaboration}}]{2000AJ....120.1579Y}
{York}, D.~G., {Adelman}, J., {Anderson}, John~E., J., {et~al.} 2000, \aj, 120,
  1579, \dodoi{10.1086/301513}

\bibitem[{{Yuan} {et~al.}(2013){Yuan}, {Zhang}, {Zhang}, {Lei}, {Dong}, \&
  {Zhao}}]{2013A&C.....3...65Y}
{Yuan}, H., {Zhang}, H., {Zhang}, Y., {et~al.} 2013, Astronomy and Computing,
  3, 65, \dodoi{10.1016/j.ascom.2013.12.001}

\end{thebibliography}
\bibliographystyle{aasjournal}



\end{document}